\documentclass[aps,groupedaddress,floatfix,nofootinbib,preprintnumbers,eqsecnum]{revtex4}
\usepackage{amssymb,amsmath,graphicx,multirow}

\begin{document}

\preprint{UH511-1197-12}

\title{Direct Detection of Dynamical Dark Matter}
\author{Keith R. Dienes$^{1,2,3}$\footnote{E-mail address:  {\tt dienes@physics.arizona.edu}},
      Jason Kumar$^{4}$\footnote{E-mail address:  {\tt jkumar@phys.hawaii.edu}},  
      Brooks Thomas$^{4}$\footnote{E-mail address:  {\tt thomasbd@phys.hawaii.edu}}}
\affiliation{
     $^1$ Physics Division, National Science Foundation, Arlington, VA  22230  USA\\
     $^2$ Department of Physics, University of Maryland, College Park, MD  20742  USA\\
     $^3$ Department of Physics, University of Arizona, Tucson, AZ  85721  USA\\
     $^4$ Department of Physics, University of Hawaii, Honolulu, HI 96822  USA}

\begin{abstract}
Dynamical dark matter (DDM) is an alternative framework for dark-matter 
physics in which the dark-matter candidate is an ensemble of constituent 
fields with differing masses, lifetimes, and cosmological abundances.  In 
this framework, it is the balancing of these quantities against each other 
across the ensemble as a whole which ensures phenomenological viability. 
In this paper, we examine the prospects for the direct detection of a DDM 
ensemble.  In particular, we study the constraints imposed by current 
limits from direct-detection experiments on the parameter space of DDM 
models, and we assess the prospects for detecting such an ensemble and 
distinguishing it from traditional dark-matter candidates on the basis of 
data from the next generation of direct-detection experiments.  For 
concreteness, we focus primarily on the case in which elastic scattering 
via spin-independent interactions dominates the interaction rate between 
atomic nuclei and the constituent particles of the ensemble.  We also 
briefly discuss the effects of modifying these assumptions. 
\end{abstract}

\maketitle

\newcommand{\newc}{\newcommand}
\newc{\gsim}{\lower.7ex\hbox{$\;\stackrel{\textstyle>}{\sim}\;$}}
\newc{\lsim}{\lower.7ex\hbox{$\;\stackrel{\textstyle<}{\sim}\;$}}
\makeatletter
\newcommand{\biggg}{\bBigg@{3}}
\newcommand{\Biggg}{\bBigg@{4}}
\makeatother

\def\vac#1{{\bf \{{#1}\}}}

\def\beq{\begin{equation}}
\def\eeq{\end{equation}}
\def\beqn{\begin{eqnarray}}
\def\eeqn{\end{eqnarray}}
\def\calM{{\cal M}}
\def\calV{{\cal V}}
\def\calF{{\cal F}}
\def\half{{\textstyle{1\over 2}}}
\def\quarter{{\textstyle{1\over 4}}}
\def\ie{{\it i.e.}\/}
\def\eg{{\it e.g.}\/}
\def\etc{{\it etc}.\/}


\def\inbar{\,\vrule height1.5ex width.4pt depth0pt}
\def\IR{\relax{\rm I\kern-.18em R}}
 \font\cmss=cmss10 \font\cmsss=cmss10 at 7pt
\def\IQ{\relax{\rm I\kern-.18em Q}}
\def\IZ{\relax\ifmmode\mathchoice
 {\hbox{\cmss Z\kern-.4em Z}}{\hbox{\cmss Z\kern-.4em Z}}
 {\lower.9pt\hbox{\cmsss Z\kern-.4em Z}}
 {\lower1.2pt\hbox{\cmsss Z\kern-.4em Z}}\else{\cmss Z\kern-.4em Z}\fi}
\def\thbar{\bar{\theta}}
\def\fhatPQ{\hat{f}_{\mathrm{PQ}}}
\def\fPQ{f_{\mathrm{PQ}}}
\def\mPQ{m_{\mathrm{PQ}}}
\def\wtl{\widetilde{\lambda}}
\def\ta{\widetilde{a}}
\def\TBBN{T_{\mathrm{BBN}}}
\def\OmegaCDM{\Omega_{\mathrm{CDM}}}
\def\OmegaDM{\Omega_{\mathrm{CDM}}}
\def\Omegatot{\Omega_{\mathrm{tot}}}
\def\rhocrit{\rho_{\mathrm{crit}}}
\def\alMRE{a_{\lambda,\mathrm{M}}}
\def\aldotMRE{\dot{a}_{\lambda,\mathrm{M}}}
\def\tMRE{t_{\mathrm{MRE}}}
\def\TMRE{T_{\mathrm{MRE}}}
\def\tQCD{t_{\mathrm{QCD}}}
\def\tauMRE{\tau_{\mathrm{M}}}
\def\mPQdot{\dot{m}_{\mathrm{PQ}}}
\def\mPQddot{\ddot{m}_{\mathrm{PQ}}}
\def\mPQbar{\overline{m}_{\mathrm{PQ}}}
\def\mXdot{\dot{m}_X}
\def\mXddot{\ddot{m}_X}
\def\mXbar{\overline{m}_X}
\def\TRH{T_{\mathrm{RH}}}
\def\tRH{t_{\mathrm{RH}}}
\def\LambdaQCD{\Lambda_{\mathrm{QCD}}}
\def\fhatX{\hat{f}_X}
\def\tnow{t_{\mathrm{now}}}
\def\Omvac{\Omega_{\mathrm{vac}}^{(0)}}
\def\arcsinh{\mbox{arcsinh}}
\def\zRH{z_{\mathrm{RH}}}
\def\zMRE{z_{\mathrm{MRE}}}
\def\zinit{z_{\mathrm{init}}}
\def\tinit{t_{\mathrm{init}}}
\def\sinit{s_{\mathrm{init}}}
\def\sRH{s_{\mathrm{RH}}}
\def\sMRE{s_{\mathrm{MRE}}}
\def\snow{s_{\mathrm{now}}}
\def\BRgamma{\mathrm{BR}_{\lambda}^{(2\gamma)}}
\def\te{t_{\mathrm{early}}}
\def\tl{t_{\mathrm{late}}}
\def\Ehi{E_{\mathrm{high}}}
\def\Elo{E_{\mathrm{low}}}
\def\tBBN{t_{\mathrm{BBN}}}
\def\tosc{t_{\mathrm{osc}}}
\def\Tosc{T_{\mathrm{osc}}}
\def\Tnow{T_{\mathrm{now}}}
\def\Tmax{T_{\mathrm{max}}}
\def\fhatX{\hat{f}_X}
\def\LambdaG{\Lambda_G}
\def\mX{m_X}
\def\tX{t_X}
\def\tG{t_G}
\def\OmegaDM{\Omega_{\mathrm{CDM}}}
\def\ntrans{n_{\mathrm{trans}}}
\def\nosc{n_{\mathrm{osc}}}
\def\ninf{n_{\mathrm{inf}}}
\def\nG{n_{G}}
\def\ndec{n_{\mathrm{dec}}}
\def\ncut{n_{\mathrm{cut}}}
\def\nexpl{n_{\mathrm{expl}}}
\def\tLS{t_{\mathrm{LS}}}
\def\Ech{T_{\mathrm{ch}}}
\def\Omegavac{\Omega_{\mathrm{vac}}}
\def\etanow{\eta_\ast}
\def\lambdadec{\lambda_{\mathrm{dec}}}
\def\lambdatrans{\lambda_{\mathrm{trans}}}
\def\Omegatotnow{\Omega_{\mathrm{tot}}^\ast}
\def\sigmaSI{\sigma^{(\mathrm{SI})}}
\def\weff{w_{\mathrm{eff}}}
\def\vmin{v_{\mathrm{min}}}
\def\vmax{v_{\mathrm{min}}}
\def\vesc{v_{\mathrm{esc}}}
\def\erf{\mathrm{erf}}
\def\Emin{E_R^{\mathrm{min}}}
\def\Emax{E_R^{\mathrm{max}}}
\def\keVee{\mathrm{keV}_{\mathrm{ee}}}
\def\keVnr{\mathrm{keV}_{\mathrm{nr}}}
\def\Leff{\mathcal{L}_{\mathrm{eff}}}
\def\Ivint{\mathcal{I}(m_j)}
\def\rhototloc{\rho^{\mathrm{loc}}_{\mathrm{tot}}}


\input epsf





\section{Introduction\label{sec:intro}}


Dynamical dark matter (DDM)~\cite{DynamicalDM1,DynamicalDM2} has recently been 
advanced as an alternative framework for dark-matter physics.  In this framework, 
the usual assumption of dark-matter stability is replaced by a balancing between 
lifetimes and cosmological abundances across a vast ensemble of particles which 
collectively constitute the dark matter.  Within this framework, the 
dark-matter candidate is the full ensemble itself --- a collective entity which 
cannot be characterized in terms of a single, well-defined mass, lifetime, 
or set of interaction cross-sections 
with visible matter.   As a result, cosmological quantities such as the total 
relic abundance $\Omegatot$ of the ensemble, its composition, and its equation
of state are time-dependent (\ie, dynamical) and evolve throughout the history of 
the universe.  Moreover, for this same reason, DDM ensembles also give rise to a variety of 
distinctive experimental signatures which serve to distinguish them from 
traditional dark-matter candidates.  A number of phenomenological and 
cosmological consequences to which DDM ensembles can give rise were presented 
in Refs.~\cite{DynamicalDM2,DynamicalDM3}, along 
with the bounds such effects imply on the parameter space of an explicit 
model within the general DDM framework.
DDM ensembles can also give rise to characteristic signatures at 
colliders~\cite{DDMColliders}, including distinctive imprints on the 
kinematic distributions of the Standard-Model (SM) particles produced in
conjunction with the dark-sector fields.   
   
In this paper, we examine the prospects for the direct detection of 
DDM ensembles via their interactions with atomic nuclei --- a detection
strategy~\cite{GoodmanWitten} which has come to play an increasingly 
central role in the phenomenology of most proposed dark-matter candidates
(for reviews, see, \eg, Ref.~\cite{DirectDetReviews}).  
Indeed, conclusive evidence of nuclear recoils induced by the 
scattering of particles in the dark-matter halo would provide the most 
unambiguous and compelling signal --- and moreover the only 
non-gravitational evidence --- for particle dark matter to date.  
Data from the current generation of direct-detection 
experiments have already placed stringent constraints on many models of the 
dark sector, and the detection prospects have been investigated for a variety 
of traditional dark-matter candidates at next generation of such experiments.
Studies in the context of particular multi-component models of the 
dark sector have also been 
performed~\cite{MultiComponentBlock,ProfumoTwoComponentDirectDet}.   

Here, we shall demonstrate that DDM ensembles can give rise to 
distinctive features in the recoil-energy spectra observed at direct-detection 
experiments, and that these features can serve to distinguish DDM ensembles 
from traditional dark-matter candidates.  These features 
include resolvable kinks in the recoil-energy spectra, as well as 
characteristic shapes which are difficult to realize within the context of 
traditional models --- particularly under standard astrophysical assumptions.  
As we shall demonstrate, these features should be distinguishable for a 
broad range of DDM scenarios at the next generation of direct-detection 
experiments.  Of course, the potential for differentiation within the appropriate 
limiting regimes of DDM parameter space accords with those obtained in
previous studies of two-component models~\cite{ProfumoTwoComponentDirectDet}. 
However, as we shall demonstrate, the assumption of a full DDM ensemble
as our dark-matter candidate leads to many distinctive features which
emerge in significant regions of parameter space and which
transcend those which arise for models with only a few dark-sector particles.

This paper is organized as follows.
In Sect.~\ref{sec:ScatteringTradDM}, we review the general aspects of 
dark-matter direct detection.  We discuss how considerations related to 
particle physics, nuclear physics, and astrophysics impact both the 
differential and total rate for the inelastic scattering of dark-matter 
particles with atomic nuclei, and examine the properties of the 
recoil-energy spectra associated with traditional dark-matter candidates.   
In Sect.~\ref{sec:ScatteringDDM}, by contrast, we investigate 
how these results are modified when the dark-matter candidate is 
a DDM ensemble, and we compare the resulting recoil-energy spectra to those obtained in 
traditional dark-matter models.  In Sect.~\ref{sec:Limits}, we derive a set of constraints 
on the parameter space of DDM models from current direct-detection data, and in
Sect.~\ref{sec:Prospects}, we discuss the prospects for obtaining evidence of 
DDM ensembles at the next generation of direct-detection experiments and 
for distinguishing such ensembles from traditional dark-matter candidates.   
Finally, in Sect.~\ref{sec:Conclusions}, we summarize our results and discuss 
possible directions for future study.  


\section{Direct Detection: Preliminaries\label{sec:ScatteringTradDM}}


We begin our study by briefly reviewing the situation in which a 
traditional dark-matter candidate $\chi$ with a mass $m_\chi$ scatters 
off a collection of atomic nuclei.
In general, the differential rate (per unit mass of detector material) for 
the scattering of such a dark-matter candidate off 
a collection of atomic nuclei can be written in the form~\cite{DirectDetReviews}
\begin{equation}
  \frac{dR}{dE_R}~=~\frac{\sigma_{N\chi}^{(0)}\rho_\chi^{\mathrm{loc}}}
     {2m_\chi\mu^2_{N\chi}}F^2(E_R) I_\chi(E_R)~,
  \label{eq:DiffRateStandardDM} 
\end{equation}
where $E_R$ is the recoil energy of the scattered nucleus in the reference frame
of the detector, $\sigma_{N\chi}^{(0)}$ is the $\chi$-nucleus scattering 
cross-section at zero momentum transfer, $\rho_\chi^{\mathrm{loc}}$ is 
the {\it local}\/ energy density of $\chi$,
$F(E_R)$ is a nuclear form factor, $m_N$ is the mass of the scattered nucleus $N$,
$\mu_{N\chi}\equiv m_\chi m_N/(m_\chi+m_N)$ is the 
reduced mass of the $\chi$-nucleus system, and $I_\chi(E_R)$ is the
mean inverse speed of $\chi$ in the dark-matter halo for a given $E_R$.  This
mean inverse speed, which encodes the relevant information about the halo-velocity 
distribution of $\chi$, is given by  
\begin{equation}
  I_\chi(E_R) ~\equiv~ \int_{v>\vmin}\frac{\mathcal{F}_\chi(\vec{v})}{v}d^3v~,
  \label{eq:DefOfIChi}
\end{equation}
where $\mathcal{F}_\chi(\vec{v})$ denotes the distribution of the 
detector-frame velocities $\vec{v}$ of the $\chi$ in the local dark-matter 
halo and where $v\equiv|\vec{v}|$.    
The lower limit $\vmin$ on $v$ follows from the
condition that only those $\chi$ with velocities in excess of the kinematic 
threshold for non-relativistic scattering 
\begin{equation}
  \vmin ~\equiv~ \sqrt{\frac{E_Rm_N}{2\mu_{N\chi}^2}}~, 
  \label{eq:vminElasticStdDM}
\end{equation}  
can contribute to the scattering rate.  Moreover, the halo-velocity distribution
$\mathcal{F}_\chi(\vec{v})$ itself is truncated at 
$|\vec{v} + \vec{v}_e| < \vesc$, where $\vec{v}_e$ is the velocity of 
the Earth with respect to the dark-matter halo, and where $\vesc$ is the 
galactic escape velocity.  Indeed, any dark-matter particle with a speed in 
excess of $\vesc$ in the rest frame of the dark-matter halo would likely 
have escaped from the galaxy long ago. 

One of the primary challenges in interpreting direct-detection data is that 
substantial uncertainties exist in many of the quantities appearing
in Eq.~(\ref{eq:DiffRateStandardDM}). 
It is therefore necessary for one to make certain assumptions about the
properties of the dark-matter halo, the nuclear form factor, \etc, in 
order to make concrete predictions regarding the detection prospects 
for any given theory of dark matter.  
Consequently, in this paper, we adopt a ``standard benchmark''
set of well-motivated assumptions concerning the relevant quantities 
in Eq.~(\ref{eq:DiffRateStandardDM}).

The first class of assumptions which define our standard benchmark
are those related to particle physics.  In particular, we take the dark sector 
to comprise a traditional dark-matter particle $\chi$ which scatters purely 
elastically off nuclei.  Moreover, spin-independent scattering is assumed to 
dominate the total scattering rate.
It follows that $\sigma_{N\chi}^{(0)}$ may be written in the form 
\begin{equation}
  \sigma_{N\chi}^{(0)}~=~\frac{4\mu_{Nj}^2}{\pi}\big[Zf_{p\chi}+(A-Z)f_{n\chi}\big]^2~,
  \label{eq:GeneralSpinIndepXSecNuclei}
\end{equation} 
where $f_{p\chi}$ and $f_{n\chi}$ are the respective effective couplings of 
$\chi$ to the proton and neutron, $Z$ is the atomic number of the nucleus in 
question, and $A$ is its atomic mass.  In addition, the interactions between $\chi$ and 
nucleons are taken to be isospin-conserving, in the sense that
$f_{p\chi} = f_{n\chi}$; hence Eq.~(\ref{eq:GeneralSpinIndepXSecNuclei}) reduces to
\begin{equation}
  \sigma_{N\chi}^{(0)}~=~\frac{4\mu_{N\chi}^2}{\pi}f_{n\chi}^2 A^2~.
  \label{eq:sigmachi}
\end{equation} 
Similarly, it is also useful to define the spin-independent cross-section 
{\it per nucleon}\/ at zero-momentum transfer:    
\begin{equation}
  \sigmaSI_{n\chi}~\equiv~\frac{4\mu_{n\chi}^2}{\pi}f_{n\chi}^2
   ~=~  \sigma_{N\chi}^{(0)} \frac{\mu_{n\chi}^2}{\mu_{N\chi}^2A^2} ~,
  \label{eq:sigmaSIchi}
\end{equation}
where $\mu_{n\chi}$ is the reduced mass of the $\chi$-nucleon system.
This quantity has the advantage of being essentially independent of the 
properties of the target material, and therefore useful for comparing data
from different experiments.
  
The second class of assumptions which define our standard benchmark 
are those related to the astrophysics of the dark-matter halo.  
In particular, the local dark-matter 
density is taken to be $\rhototloc \approx 0.3~\mathrm{GeV/cm}^{3}$ and
the velocity distribution of particles in the dark-matter halo is
taken to be Maxwellian.  From the latter assumption, it follows that
the integral over halo velocities in Eq.~(\ref{eq:DefOfIChi}) 
is~\cite{LewinSmith,SavageFreese}  
\begin{equation}
  I_\chi(E_R)~=~
    \frac{k}{2v_e}\times \begin{cases}  \displaystyle
    \erf\left(\frac{\vmin + v_e}{v_0}\right)
         -\erf\left(\frac{\vmin - v_e}{v_0}\right) 
         - \frac{4v_e}{v_0\sqrt{\pi}}e^{-\vesc^2/v_0^2} & 
     \vmin \leq \vesc - v_e \\
    \rule{0cm}{0.75cm} \displaystyle
    \erf\left(\frac{\vesc}{v_0}\right)
         -\erf\left(\frac{\vmin - v_e}{v_0}\right) 
         - \frac{2(\vesc+v_e-\vmin)}{v_0\sqrt{\pi}}
         e^{-\vesc^2/v_0^2} &
     \vesc - v_e < \vmin \leq \vesc + v_e \\
     \rule{0cm}{0.75cm}
    0 & \vmin > \vesc + v_e
   \end{cases}
   \label{eq:VelocDistIntCases}
\end{equation}
where $v_0 \approx 220$~km/s is the local circular velocity and 
\begin{equation}
  k ~\equiv~ \left[\erf\left(\frac{\vesc}{v_0}\right) 
      - \frac{2\vesc}{v_0\sqrt{\pi}}e^{-\vesc^2/v_0^2}\right]^{-1}
\end{equation}
is a coefficient which is independent of both time and $m_\chi$.  
By contrast, $v_e$ is time-dependent and modulates annually due to the 
revolution of the Earth around the Sun.  However, in this paper, 
we focus primarily on the time-averaged scattering rate observed at a 
given experiment.  We therefore approximate the expression in 
Eq.~(\ref{eq:VelocDistIntCases}) by replacing $v_e$ with its annual 
average $\langle v_e\rangle \approx 1.05\, v_0$ in what follows.  
Finally, the galactic escape velocity is taken to be 
$\vesc \approx 540\mathrm{~km/s}$, in accord with the results obtained 
from the RAVE survey~\cite{RAVESurvey}.
 
The third class of assumptions which define our standard benchmark 
are those related to nuclear physics.  These assumptions are 
collectively embodied by the nuclear form factor $F(E_R)$.  In our standard 
benchmark, this form factor is taken to have the Helm 
functional form~\cite{HelmFormFactor} 
\begin{equation}
  F(E_R)~=~ \frac{3J_1(\sqrt{2m_N E_R} R_1)}{\sqrt{2 m_N E_R} R_1}\, 
  e^{-m_N E_R s^2}~,
  \label{eq:HelmFormFactor}
\end{equation}  
where $J_1(x)$ denotes the spherical Bessel function, 
$s \approx 0.9$~fm is an empirically-determined 
length scale, and $R_1 \equiv \sqrt{R^2-5s^2}$, with 
$R \equiv (1.2\mathrm{~fm})\times A^{1/3}$.  Note that  
there exist particular values of $E_R$ at which $F(E_R)$ 
vanishes in this form-factor model (due to the zeroes of the Bessel 
function), in the vicinity of which results derived using 
Eq.~(\ref{eq:HelmFormFactor}) are unreliable.  However, turns out 
that all such values of $E_R$ will lie well outside the range relevant 
for our analysis.
     
Of course, deviations from this standard benchmark can have a potentially 
significant impact on recoil-energy spectra.  For example, the effects of 
unorthodox coupling structures~\cite{IVDM,MagneticDM},
more complicated velocity distributions~\cite{NonStandardHaloVelDist},
different values of the local dark-matter energy density~\cite{Salucci},
and alternative form-factor models~\cite{FormFactorDM}
have all been investigated in the literature.  In this paper, however, 
we shall concentrate on the effects that arise when a traditional dark-matter
candidate is replaced by a DDM ensemble and hold all other aspects of our 
standard benchmark fixed.

\begin{figure}[ht!]
\centerline{
  \epsfxsize 3.0 truein \epsfbox {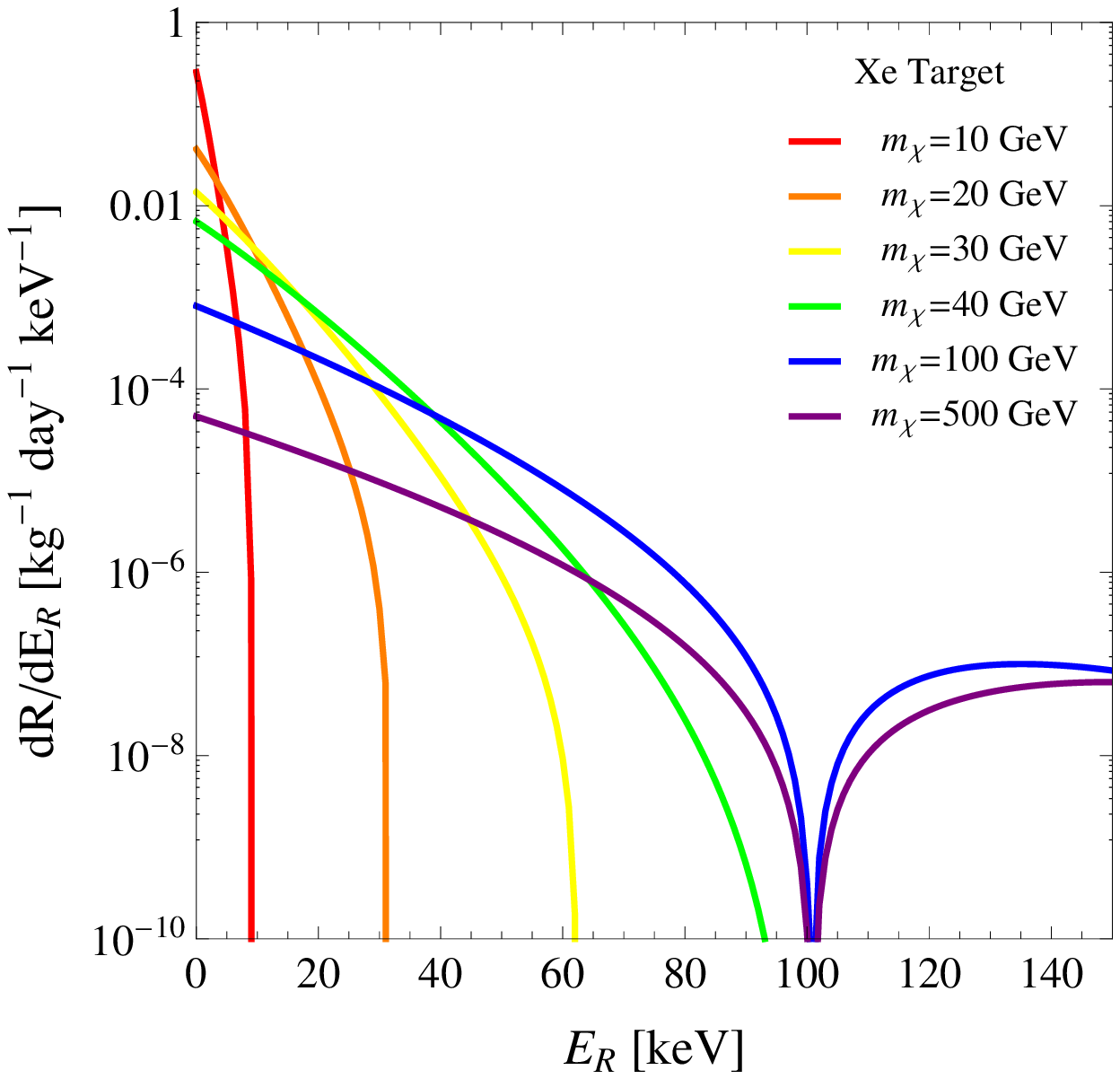} ~~~
  \epsfxsize 3.0 truein \epsfbox {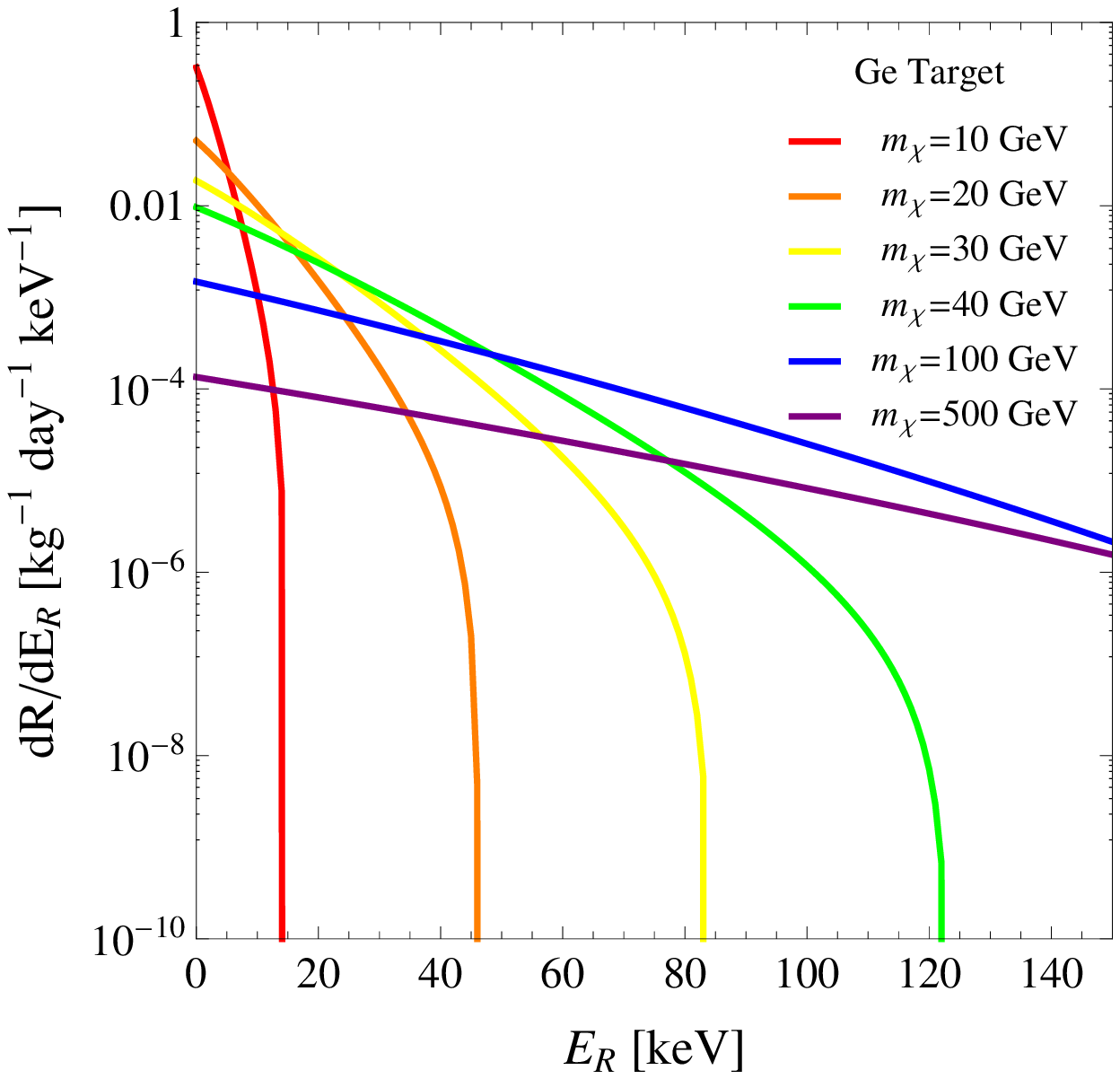} }
\caption{A representative set of recoil-energy spectra obtained for a traditional
dark-matter candidate $\chi$ with a mass $m_\chi$ scattering off a xenon target
(left panel) and a germanium target (right panel).   
\label{fig:StdDMdrdERPlots}}
\end{figure}

The standard benchmark described above already leads to characteristic 
spectra for traditional dark-matter candidates.   
In Fig.~\ref{fig:StdDMdrdERPlots}, we display a set of recoil-energy spectra
associated with the spin-independent scattering of such dark-matter candidates 
off xenon nuclei (left panel) and germanium nuclei (right panel) for our standard 
benchmark.  In each panel, the curves shown correspond to several different 
values of $m_\chi$, and each of these curves is normalized such that 
$\sigma_{N\chi}^{(0)} = 1$~pb.

The shapes of the curves shown in Fig.~\ref{fig:StdDMdrdERPlots}
are determined primarily by two physical effects which suppress the 
differential event rate at large $E_R$.  One of these effects stems 
from the distribution of particle velocities in the dark-matter halo.
Since $\mathcal{F}(\vec{v})$ falls off exponentially at velocities above
$|\vec{v} + \vec{v}_e| \sim v_0$, the recoil-energy spectra likewise 
experience a similar suppression 
above $E_R \sim 2 (v_0 + v_e)^2 \mu_{N\chi}^2/ m_N$.  Moreover, these spectra are also 
truncated at $E_R \sim 2 (\vesc + v_e)^2 \mu_{N\chi}^2/ m_N$ as a result of the 
galactic escape velocity.  These effects are particularly important for 
$m_\chi \lesssim 40$~GeV and become increasingly pronounced as $m_\chi$ decreases.
Indeed, we will see in Sect.~\ref{sec:ScatteringDDM} that these effects turn out to 
play a critical role in the direct-detection phenomenology of DDM ensembles, 
precisely because they are sensitive to $m_\chi$.  

The other effect which plays a significant role in determining the shape of the 
recoil-energy spectra shown in Fig.~\ref{fig:StdDMdrdERPlots} has its origin in 
nuclear physics.  As is evident from Eq.~(\ref{eq:HelmFormFactor}), the nuclear
form factor also suppresses the differential event rate for large $E_R$. 
This suppression is particularly acute for 
heavier nuclei such as xenon ($A \approx 131$), and considerably less so for lighter
nuclei such as germanium ($A \approx 73$).  Note, however, that this effect is
independent of $m_\chi$ and depends only on $E_R$ and the properties of the 
target material.  Note also that the dip in the recoil-energy spectra displayed
in the left panel of Fig.~\ref{fig:StdDMdrdERPlots} around $E_R \sim 100$~keV
corresponds to the first zero of $F(E_R)$ in Eq.~(\ref{eq:HelmFormFactor}).
As noted above, this turns out to lie well outside the range of $E_R$ values 
relevant for this analysis.     

We see, then, that there are qualitatively two distinct regimes which 
describe the differing behaviors of the resulting recoil-energy spectra. 
In the ``low-mass regime,'' these spectra are steeply falling and highly 
sensitive to $m_\chi$.  By contrast, in the ``high-mass regime,'' these 
curves fall more slowly and are less sensitive to $m_\chi$.  
As we shall see in Sect.~\ref{sec:Prospects}, this 
distinction will ultimately play a critical role in our analysis.


\section{Direct Detection of DDM Ensembles\label{sec:ScatteringDDM}}


Let us now examine how the results obtained in Sect.~\ref{sec:ScatteringTradDM} 
are modified in the case in which the traditional dark-matter candidate is 
replaced by a DDM ensemble, with all of the other defining characteristics of our 
standard benchmark held fixed.
As discussed in Refs.~\cite{DynamicalDM1,DynamicalDM2}, 
dynamical dark matter is a new framework for 
dark-matter physics in which the notion of stability is replaced by a 
delicate balancing between lifetimes and abundances across an ensemble of 
individual dark-matter components.  As such, this framework represents the 
most general possible dark-matter sector that can be imagined while still 
satisfying astrophysical and cosmological bounds.  Furthermore, as 
discussed in Ref.~\cite{DynamicalDM1}, dynamical dark matter arises naturally in 
certain theories involving extra spacetime 
dimensions, and also in certain limits of string 
theory.  It is therefore important to consider how the direct-detection phenomenology 
of DDM ensembles differs from that of traditional dark-matter 
candidates.  Indeed, such a study can be viewed as complementary to the collider
analysis performed in Ref.~\cite{DDMColliders}.

From a direct-detection standpoint, the most salient difference between a DDM ensemble 
and a traditional dark-matter candidate is that, by definition, 
a DDM ensemble comprises a vast number of constituent fields $\chi_j$, each with 
a mass $m_j$ and local energy density $\rho_j^{\mathrm{loc}}$.  
In general, since multiple states are present in the dark 
sector, both elastic processes of the form $\chi_j N\rightarrow \chi_j N$ 
and inelastic scattering processes of the form $\chi_j N\rightarrow \chi_k N$,
with $j\neq k$, can contribute to the total $\chi_j$-nucleon scattering rate.  
In this paper, in accord with the assumptions underlying our standard picture of 
dark-matter physics, we focus primarily on the case in which elastic scattering 
provides the dominant contribution to the total scattering rate for each 
$\chi_j$.  (This occurs generically, for example, in situations in which the 
mass-splittings between all pairs of constituent particles in the ensemble 
substantially exceed 100~keV).  In this case, each $\chi_j$ also possesses a 
well-defined effective spin-independent coupling $f_{nj}$ to nucleons, and
consequently a well-defined spin-independent cross-section per nucleon
$\sigmaSI_{n\chi}\equiv 4\mu_{nj}^2f_{nj}^2/\pi$.
The total differential event rate at a given detector is   
obtained by summing over the contributions from 
each $\chi_j$, each of which is given by an expression analogous to
Eq.~(\ref{eq:DiffRateStandardDM}).  Thus, for an arbitrary DDM ensemble, 
this total differential event rate takes the form  
\begin{equation}
  \frac{dR}{dE_R}~=~\sum_j\frac{\sigma_{Nj}^{(0)}\rho_j^{\mathrm{loc}}}
     {2m_j\mu^2_{Nj}}F^2(E_R)I_j(E_R)~,
  \label{eq:DiffRateBasicSum} 
\end{equation}
subject to the constraint 
$\sum_j\rho_j^{\mathrm{loc}} = \rho^{\mathrm{loc}}_{\mathrm{tot}}$.  Note
that the nuclear form factor depends only on $E_R$ and not on the properties
of the constituent particle $\chi_j$.  By contrast, the integral over halo 
velocities depends non-trivially on $m_j$ through the kinematic threshold 
velocity 
\begin{equation}
  \vmin^{(j)} ~\equiv~ \sqrt{\frac{E_Rm_N}{2\mu_{Nj}^2}}~. 
  \label{eq:vminElasticDDM}
\end{equation}     
Note that this remains true even in the case we consider here, in which 
the velocity distributions for all $\chi_j$ are taken to be essentially identical.

The cosmology of DDM models is principally described by two 
characteristic quantities~\cite{DynamicalDM1}.  The first of these is the 
collective (present-day) relic abundance $\Omegatot$ of the full DDM ensemble, 
which is simply a sum of the individual abundances $\Omega_j$ of the $\chi_j$.  
The second quantity is 
\begin{equation}
  \eta ~\equiv~ 1 - \frac{\Omega_0}{\Omegatot}
\end{equation}
 where $\Omega_0 \equiv \max\{\Omega_j\}$; this helps to characterize the 
distribution of the $\Omega_j$ across the ensemble, and in 
particular represents the fraction of $\Omegatot$ collectively provided by all 
but the most abundant constituent.  Thus $\eta = 0$ effectively corresponds to the 
case of a traditional dark-matter candidate, while $\eta \sim \mathcal{O}(1)$ 
indicates that the full ensemble is contributing non-trivially to $\Omegatot$.

In general, the {\it local} energy densities $\rho_j^{\mathrm{loc}}$ of the 
$\chi_j$ --- which play a crucial role in direct detection --- need not have any relation 
to their {\it cosmological} abundances $\Omega_j$.  However, in typical cosmological models, 
the local energy density of any particular $\chi_j$ 
is approximately proportional to its cosmological abundance --- \ie,   
$\rho_j^{\mathrm{loc}}/\rho_{\mathrm{tot}}^{\mathrm{loc}} ~\approx~ \Omega_j/\Omegatot$.
Furthermore, we assume that $\Omegatot \approx \OmegaCDM$, so that the DDM ensemble 
contributes essentially the entirety of the cold-dark-matter relic abundance
$\OmegaCDM h^2 \approx 0.1131 \pm 0.0034$ determined by WMAP~\cite{WMAP}. 
Under these assumptions, the differential event rate in  
Eq.~(\ref{eq:DiffRateBasicSum}) may be written in the form
\begin{equation}
  \frac{dR}{dE_R}~=~
    \frac{2f_{n0}^2\rho_{\mathrm{tot}}^{\mathrm{loc}}A^2}
    {\pi m_0}
     (1-\eta)F^2(E_R)\,
     \sum_j \left(
     \frac{\Omega_j m_0 f_{nj}^2}{\Omega_0 m_j f_{n0}^2}\right)
     I_j(E_R)~, 
  \label{eq:DiffRateTotalGeneral}
\end{equation} 
where $m_0$ and $f_{n0}$ respectively denote the mass and effective coupling 
coefficient of the most abundant state $\chi_0$ in the ensemble.

For concreteness, we examine the direct-detection phenomenology of DDM 
ensembles in the context of a simplified DDM model.  In this model
$\chi_0$ is identified with the lightest state in the ensemble, and the
mass spectrum of the $\chi_j$ takes the form 
\begin{equation}
  m_j ~=~ m_0 + j^\delta \Delta m
  \label{eq:MassSpectrumForm}
\end{equation}
with $\Delta m > 0$ and $\delta > 0$, so that the $\chi_j$ are labeled in order
of increasing mass.  Moreover, in this model $\Omega_j$ and 
$f_{nj}$ are each assumed to exhibit power-law scaling with $m_j$ 
across the ensemble, so that these quantities may be written in the form
\begin{eqnarray}
  \Omega_j&=&\Omega_0\left(\frac{m_j}{m_0}\right)^{\alpha}\nonumber\\
  f_{nj}&=&f_{n0}\left(\frac{m_j}{m_0}\right)^{\beta}~,
  \label{eq:ScalingRels}
\end{eqnarray}
where $\alpha$ and $\beta$ are general power-law exponents.
Note that scaling relations of this
form emerge naturally in many realistic DDM 
scenarios~\cite{DynamicalDM1,DynamicalDM2,DDMColliders}. 
Also note that the direct-detection phenomenology of DDM ensembles 
depends on the present-day values of the $\Omega_j$ and not how these 
values have evolved in the past.  Thus the decay widths of the $\chi_j$,
although crucial for the balancing of lifetimes against abundances within
the DDM framework~\cite{DynamicalDM1}, play no role in direct detection.   

For the purposes of direct detection,
our simplified DDM ensemble is therefore characterized by two groups of parameters:
those (namely $m_0$, $\Omega_0$, and $f_{n0}$) which describe the 
properties of the most abundant state in the ensemble and which would also be 
necessary in any traditional dark-matter model, and those
(namely $\Delta m$ and the scaling exponents
$\alpha$, $\beta$, and $\delta$) which
describe how this information extends throughout the
entire ensemble.  This is therefore a very compact 
yet flexible formalism for exploring the ramifications
of having an entire DDM ensemble as our dark-matter candidate.  
However, the WMAP constraint on $\Omegatot$ fixes one of these parameters 
(most conveniently $\Omega_0$).  
Thus, the 
recoil-energy spectra to which our simplified DDM model gives rise are 
completely determined by 
$\alpha$, $\beta$, $\delta$, $m_0$, $\Delta m$, and $f_{n0}$ (or equivalently 
$\sigmaSI_{n0}$).  Note also that
the last of these parameters determines the normalization of the 
recoil-energy spectrum (\ie, the total event rate), but has no effect on the 
{\it shape}\/ of that spectrum.  

Given the scaling relations in Eqs.~(\ref{eq:MassSpectrumForm}) and~(\ref{eq:ScalingRels}),               
it is straightforward to rewrite the expressions for $\Omegatot$ and $\eta$, as well as 
the differential event rate $dR/dE_R$, in terms of these parameters.  
Indeed, in the context of our simplified DDM model, the present-day values of 
$\Omegatot$ and $\eta$ are given by   
\begin{eqnarray}
  \Omegatot &=& \Omega_0\sum_{j} 
   \left(1 + j^\delta\frac{\Delta m}{m_0}\right)^{\alpha}\nonumber \\
   \eta &=& 1 - \left[\sum_{j}
        \left(1 + j^\delta\frac{\Delta m}{m_0}\right)^{\alpha}\right]^{-1}~. 
  \label{eq:OmegatotAndEtaRel}
\end{eqnarray}
From a DDM perspective, our primary interest is in situations in which the 
number of constituent particles in the dark-matter ensemble is taken to be large.  
For this reason, we restrict our discussion to cases in which the sums in 
Eq.~(\ref{eq:OmegatotAndEtaRel}) are convergent even in the limit in which 
$j\rightarrow\infty$.  Imposing this requirement 
restricts the purview of our analysis to cases in which the condition 
$\alpha\delta < -1$ is satisfied.  Likewise,
the expression in Eq.~(\ref{eq:DiffRateTotalGeneral}) for the 
differential event rate reduces to
\begin{equation}
  \frac{dR}{dE_R}~=~
    \frac{2f_{n0}^2\rho_{\mathrm{tot}}^{\mathrm{loc}}A^2}
    {\pi m_0}
     (1-\eta)F^2(E_R)\,
     \sum_j I_j(E_R)
     \left(1+j^\delta\frac{\Delta m}{m_0}\right)^{\alpha+2\beta-1}~. 
  \label{eq:DiffRateTotalToyModel}
\end{equation} 
 
\begin{figure}[ht!]
\begin{center}
  \epsfxsize 2.25 truein \epsfbox {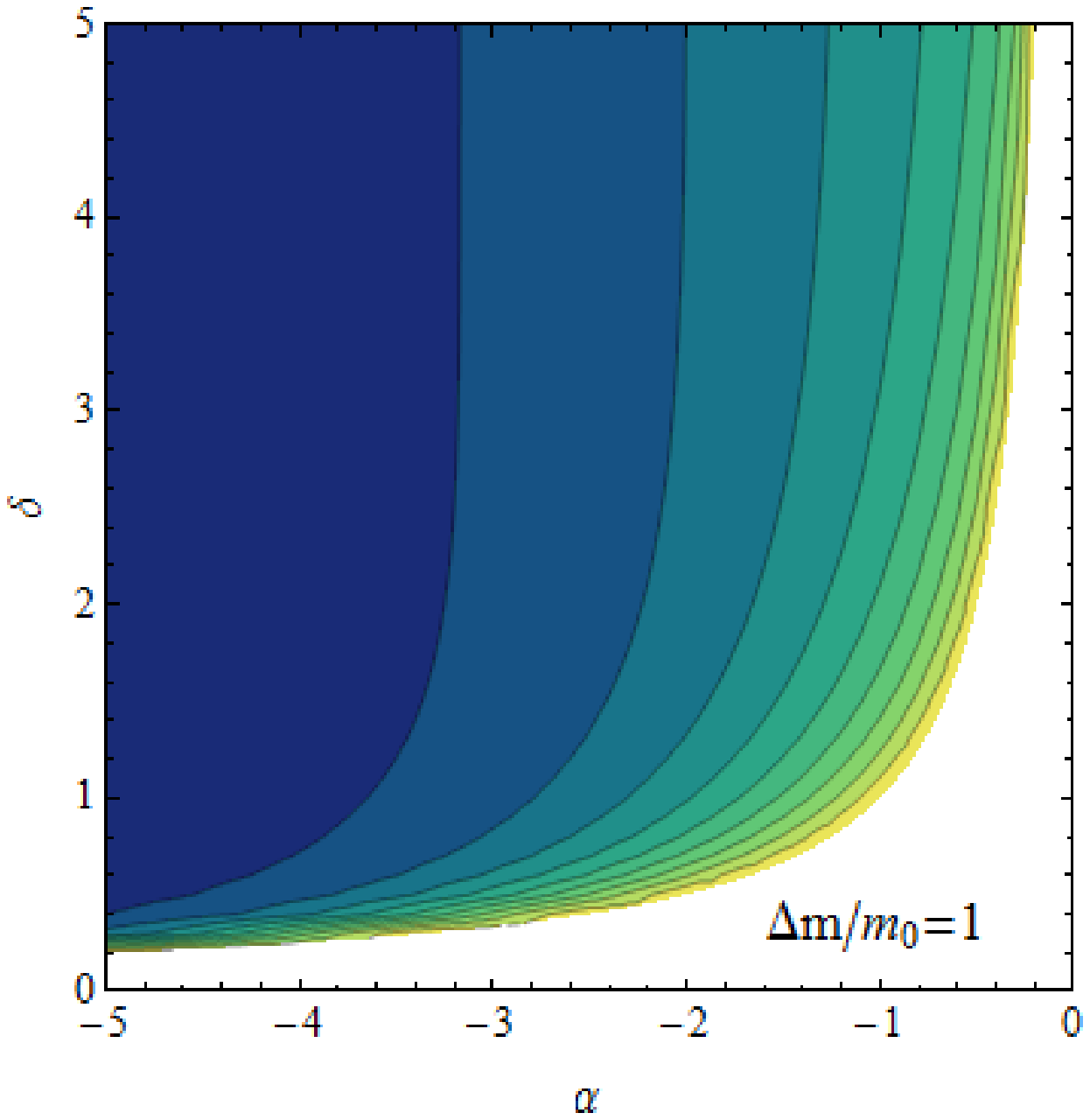}
  \epsfxsize 2.25 truein \epsfbox {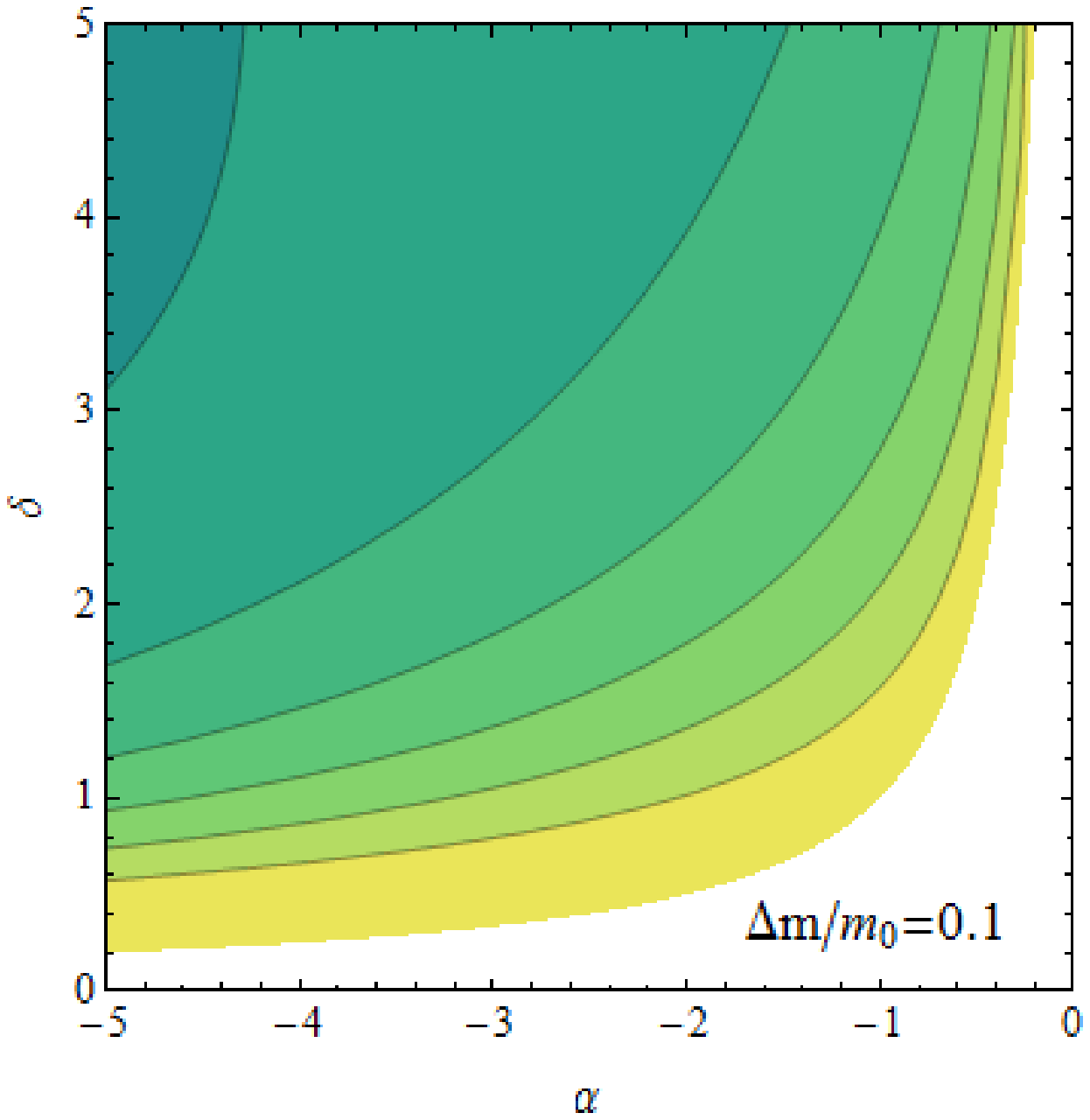}
  \epsfxsize 2.25 truein \epsfbox {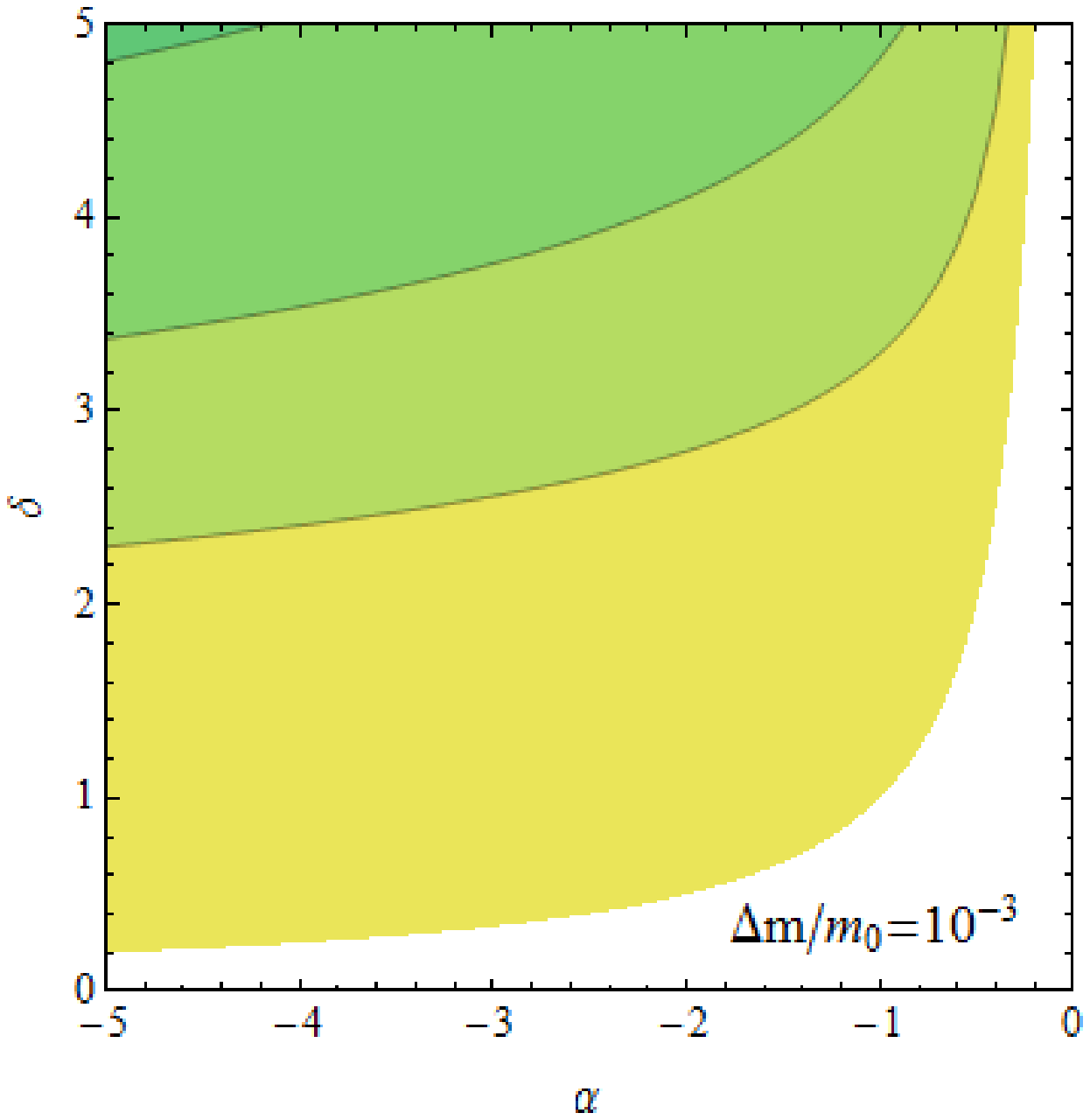}\\
  \raisebox{0.5cm}{\large $\eta$~:~~}
     \epsfxsize 5.00 truein \epsfbox {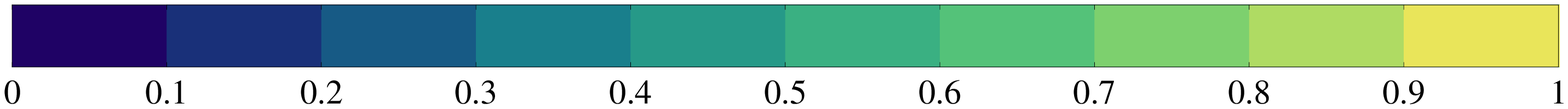} 
\end{center}
\caption{Contours of $\eta$ as a function of the
scaling coefficients $\alpha$ and $\delta$, derived under the assumption that 
$\Omegatot \approx \OmegaCDM$.  The left, center, and right panels show
 results for $\Delta m/m_0 = \{1, 0.1,10^{-3}\}$, respectively.  Note that as
$\delta m / m_0\rightarrow 0$ for fixed $\alpha$ and $\delta$, we see that 
$\eta\rightarrow 1$ and the full ensemble provides a increasingly significant 
contribution to $\Omegatot$.    
\label{fig:EtaPanels}}
\end{figure}

In Fig.~\ref{fig:EtaPanels}, we provide a series of contour plots illustrating 
the dependence of $\eta$ on the scaling coefficients $\alpha$ and $\delta$ in our 
simplified DDM model.  The left, center, and right panels of this figure display 
results for $\Delta m/m_0 = \{1, 0.1,10^{-3}\}$, respectively.  The
white region appearing in each plot is excluded by the condition 
$\alpha\delta < -1$.  The qualitative results displayed in this figure accord 
with basic intuition: $\eta$ is maximized for values of $\alpha$ and $\delta$ 
which come close to saturating the constraint $\alpha\delta < -1$, and
smaller values of the ratio $\Delta m/m_0$ for fixed $\alpha$ and $\delta$
yield larger values of $\eta$.  However, the quantitative results displayed in
Fig.~\ref{fig:EtaPanels} are less intuitive and quite significant.  In particular, 
we see that $\eta \sim \mathcal{O}(1)$ over a broad range of $\alpha$ 
and $\delta$ values, even in cases in which $\Delta m \sim m_0$.  Within this 
region of parameter space, the full DDM ensemble contributes non-trivially to 
$\Omegatot$.   

\begin{figure}[ht!]
\centerline{
  \epsfxsize 2.25 truein \epsfbox {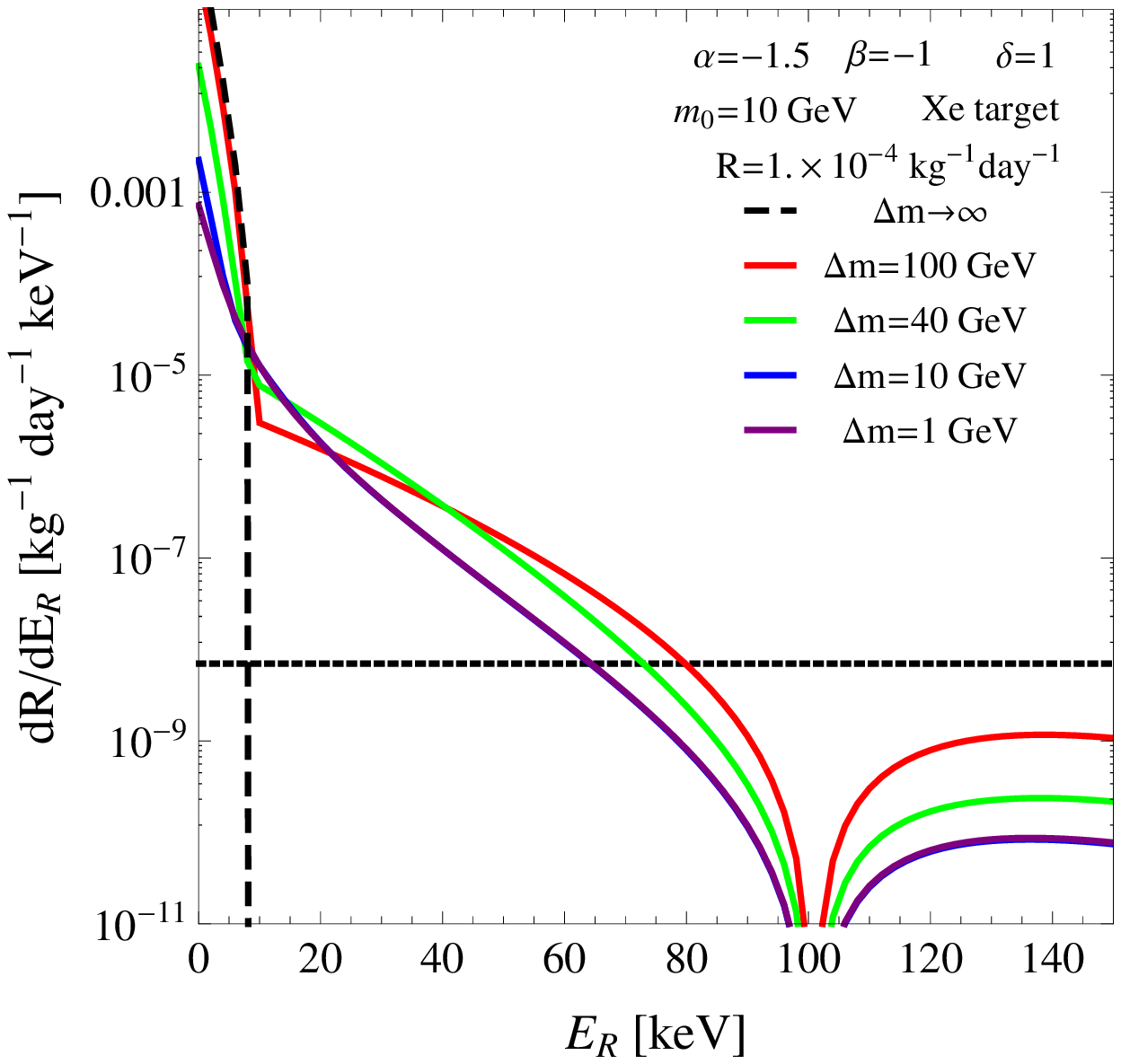}
  \epsfxsize 2.25 truein \epsfbox {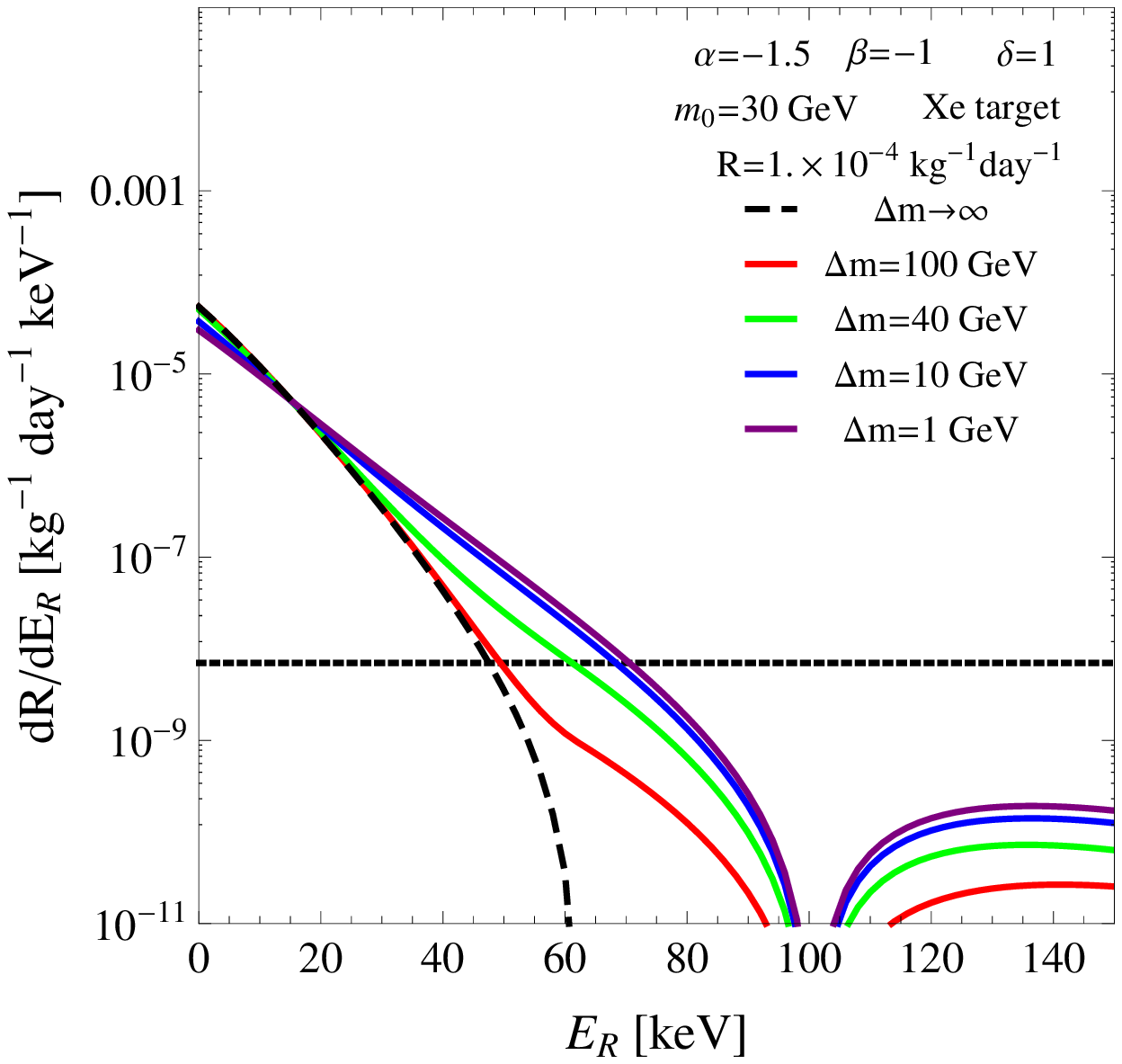}
  \epsfxsize 2.25 truein \epsfbox {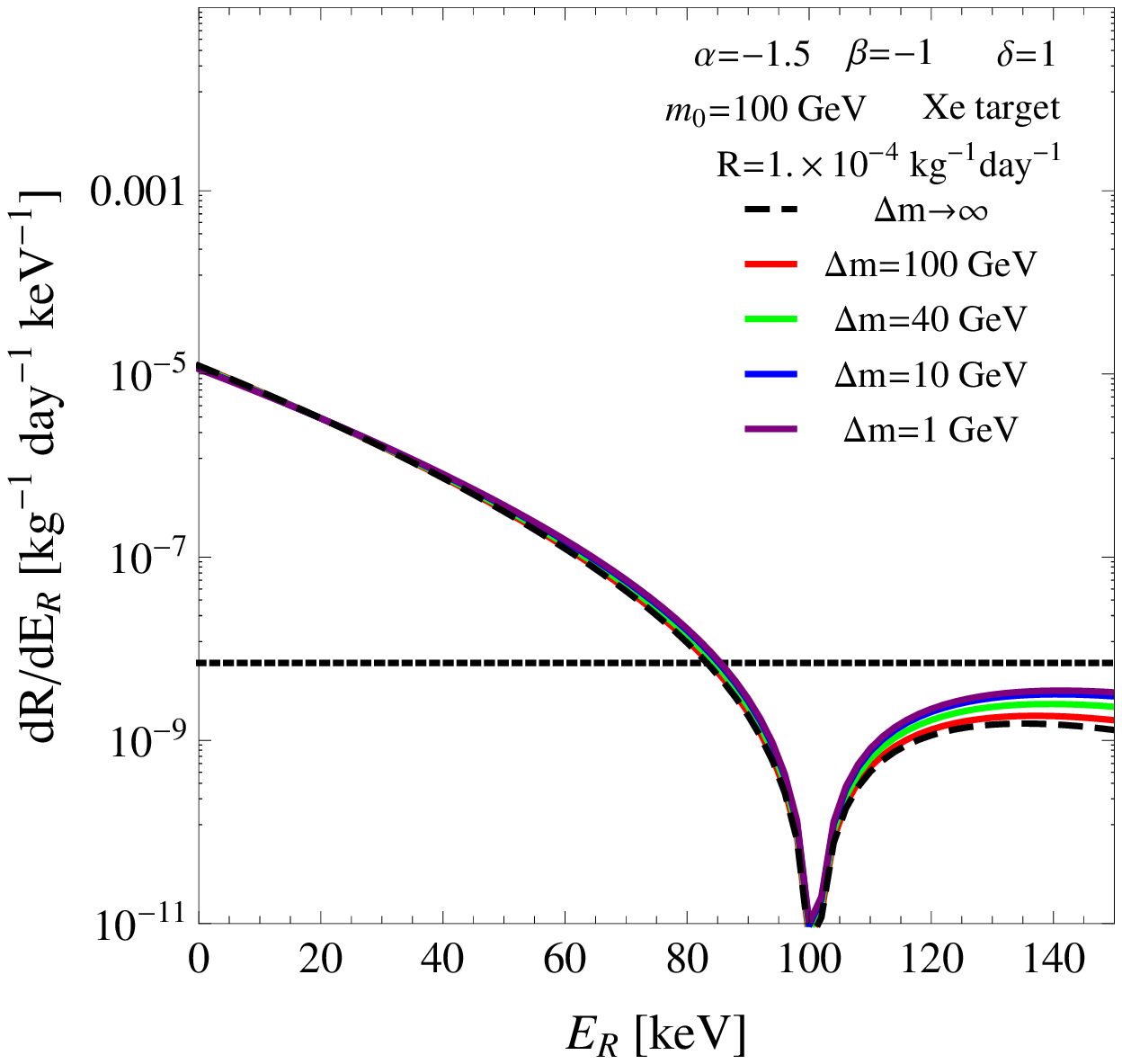}}
\caption{Recoil-energy spectra associated with DDM ensembles 
scattering elastically off of a xenon target.  In each of the three panels shown, we have 
set $\alpha = -1.5$, $\beta = -1$, and $\delta = 1$, while the left, center, and right panels
correspond to the choices $m_0 = \{10,30,100\}$~GeV, respectively.  The different curves
displayed in each panel correspond to different values of $\Delta m$, and each of these
curves has been normalized so that the total event rate for nuclear recoils in the energy 
range $8\mathrm{~keV}\lesssim E_R \lesssim 48\mathrm{~keV}$ lies just below the current
bound from XENON100 data.  Note that the 
$\Delta m \rightarrow \infty$ limit indicated by the dashed black curve corresponds to 
a traditional dark-matter candidate with a mass $m_\chi = m_0$.  The dotted black 
horizontal line indicates a reasonable estimate of the recoil-energy spectrum for
background events at the next generation of liquid-xenon detectors.
\label{fig:dRdERPanelsXe}}
\centerline{
  \epsfxsize 2.25 truein \epsfbox {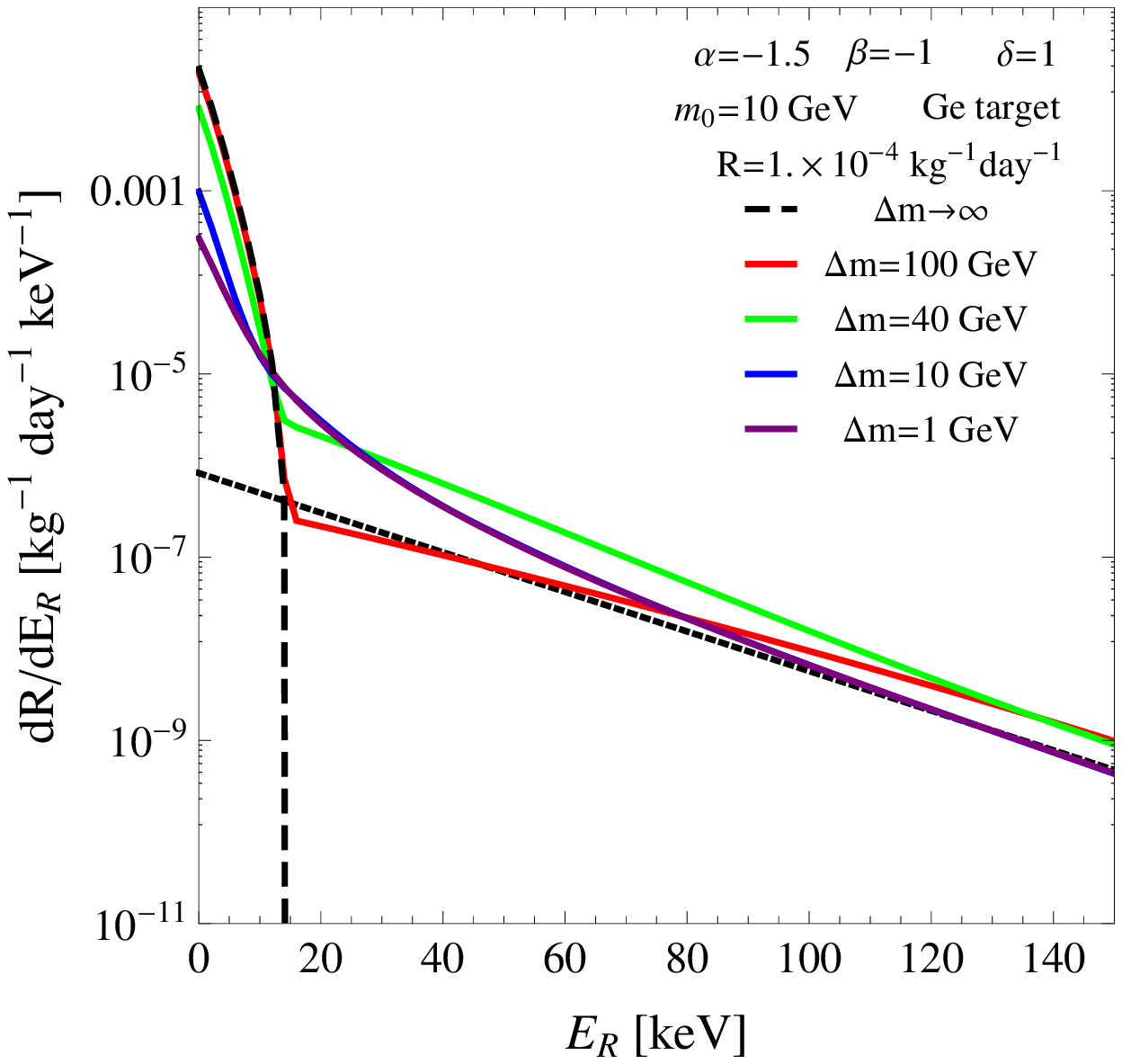}
  \epsfxsize 2.25 truein \epsfbox {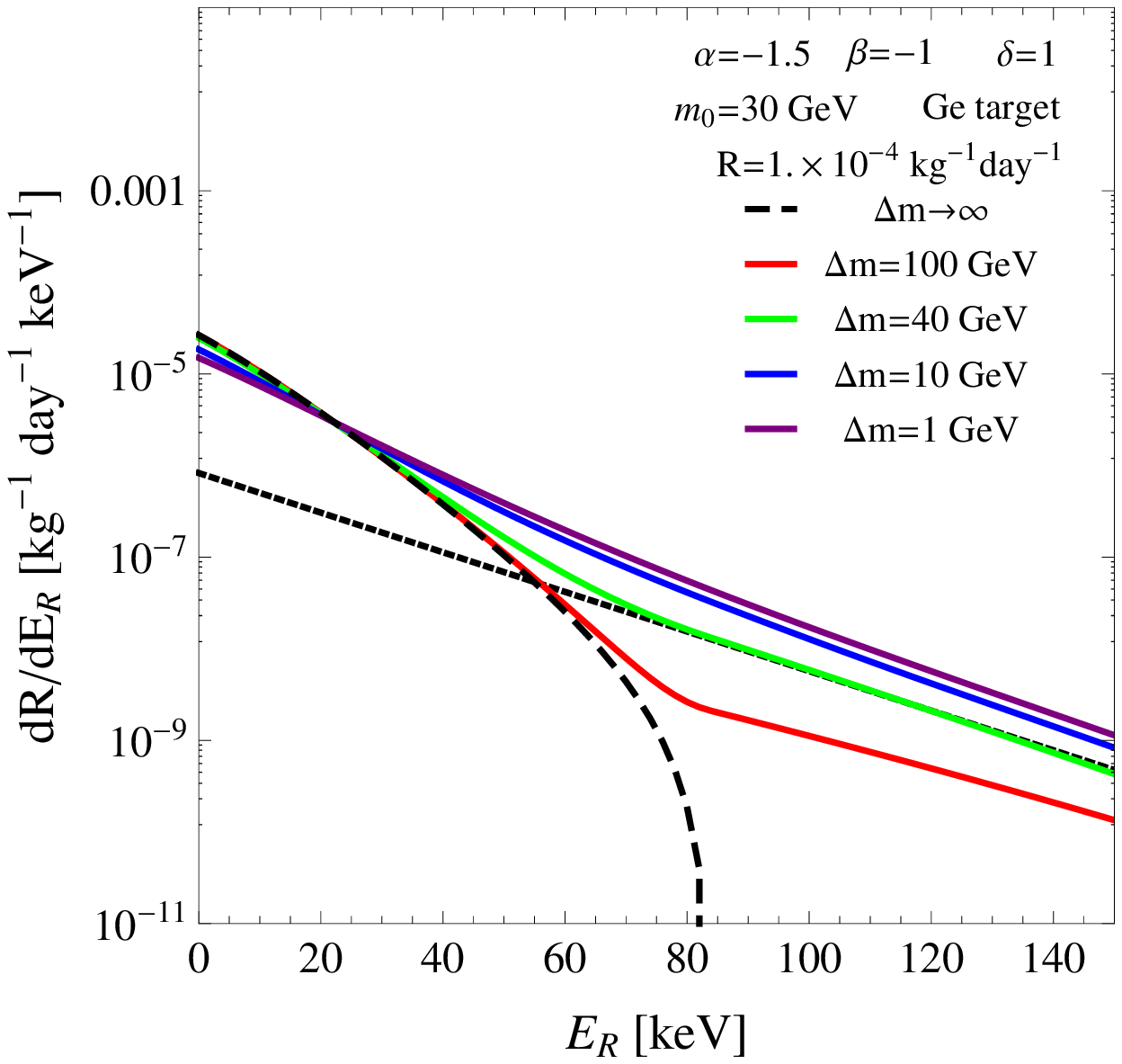}
  \epsfxsize 2.25 truein \epsfbox {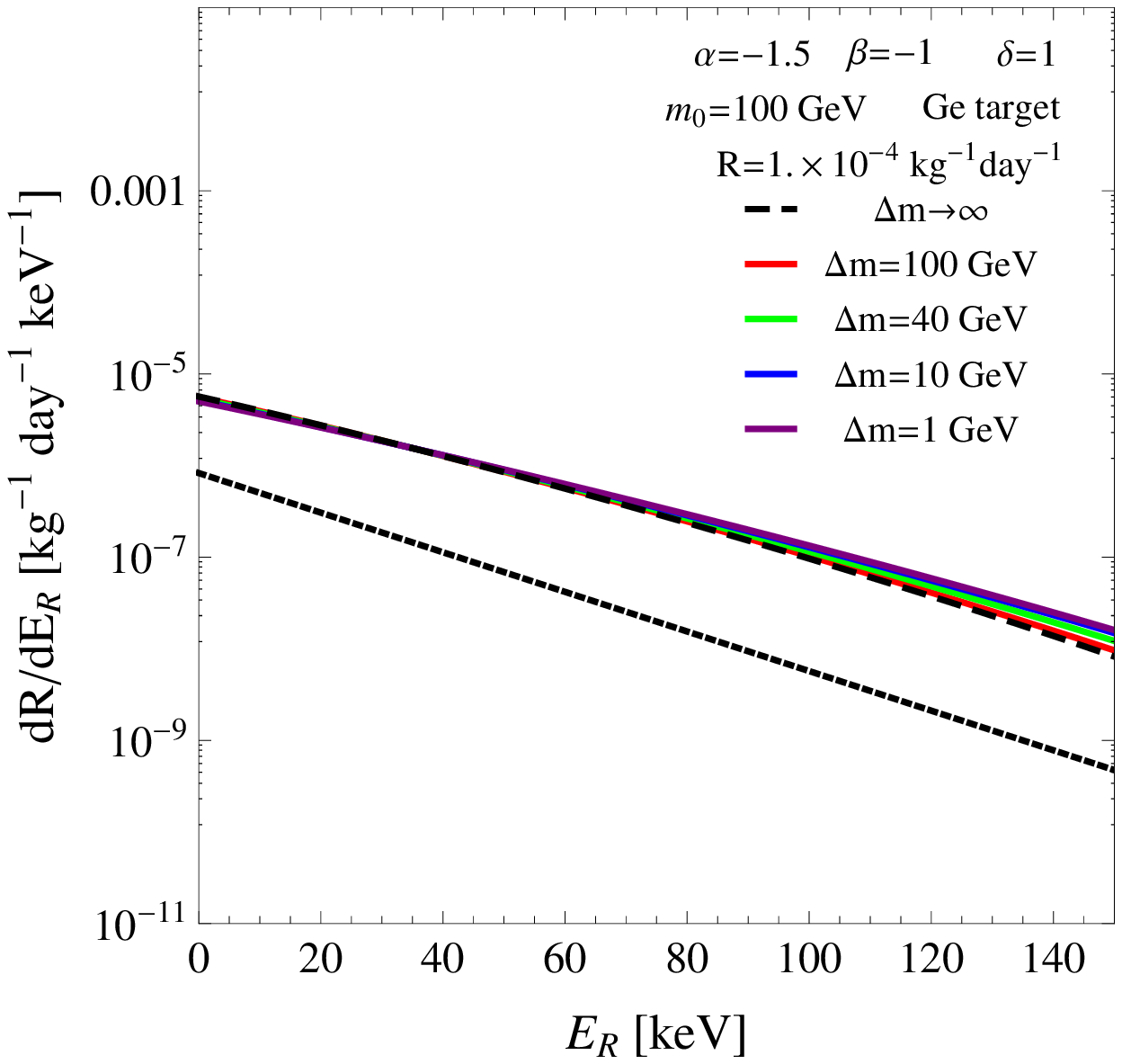}}
\caption{Recoil-energy spectra associated with DDM ensembles 
scattering elastically off of a germanium target.  The model parameters $\alpha$, $\beta$, 
and $\delta$ have been assigned as in 
Fig.~\protect\ref{fig:dRdERPanelsXe}, and we have likewise set $m_0=\{10,30,100\}$~GeV in
the left, center, and right panels of the figure, respectively.  Each of these
curves has been normalized so that the total event rate for nuclear recoils in the energy 
range $10\mathrm{~keV}\lesssim E_R \lesssim 100\mathrm{~keV}$ lies just below the current
experimental bound.  The dotted black 
horizontal line indicates a reasonable estimate of the recoil-energy spectrum for
background events at the next generation of germanium-crystal detectors.
\label{fig:dRdERPanelsGe}}
\end{figure}

We now examine the recoil-energy spectra which arise in the context of our 
simplified DDM model and identify characteristic features in these spectra.  
A representative set of such spectra is shown for
a xenon target in Fig.~\ref{fig:dRdERPanelsXe}
and for a germanium target in Fig.~\ref{fig:dRdERPanelsGe}.
In each of the three panels shown in each figure, we have set 
$\alpha = -1.5$, $\beta = -1$, and $\delta = 1$, while the left, center, and right panels
in each figure correspond to $m_0 = \{10,30,100\}$~GeV, respectively.
These values have been chosen in order to illustrate the 
different effects to which DDM ensembles can give rise.
The different curves displayed in each panel correspond to different values of 
$\Delta m$.  Each of the curves displayed in Fig.~\ref{fig:dRdERPanelsXe} has been 
normalized such that the total event rate for nuclear recoils in the energy range 
$8\mathrm{~keV}\lesssim E_R \lesssim 48\mathrm{~keV}$ 
is $R = 1.0\times 10^{-4}~\mathrm{~kg}^{-1}\mathrm{day}^{-1}$.
Likewise, each of the curves displayed in Fig.~\ref{fig:dRdERPanelsGe} 
has been normalized so that this same total event rate is obtained for 
nuclear recoils in the energy range 
$10\mathrm{~keV}\lesssim E_R \lesssim 100\mathrm{~keV}$.  Note that this rate is 
consistent with current experimental limits on the total event rate for both target
materials; moreover, these recoil-energy ranges are chosen to coincide with those
typically considered at experiments based on these respective target materials.
For reference, we also include a curve (the dotted black horizontal line) 
in Fig.~\ref{fig:dRdERPanelsXe} indicating a reasonable 
estimate of the recoil-energy spectrum for background events at the next generation of 
liquid-xenon detectors (to be discussed in more detail in Sect.~\ref{sec:Prospects}).
We likewise include an analogous curve in Fig.~\ref{fig:dRdERPanelsGe} indicating a 
reasonable estimate of the background spectrum at the next generation of 
germanium-crystal detectors. 

The results displayed in Figs.~\ref{fig:dRdERPanelsXe} and~\ref{fig:dRdERPanelsGe}  
demonstrate that the recoil-energy spectra associated with DDM ensembles and those 
associated with traditional dark-matter models differ very little for large $m_0$.  
This reflects the fact that the shape of the contribution
to the recoil-energy spectrum from any individual constituent particle $\chi_j$  
is not particularly sensitive to $m_j$ for $m_j \gtrsim 40$~GeV.  Consequently, 
for $m_0 \gtrsim 40$~GeV, the contributions from all of the $\chi_j$ in the ensemble
manifest roughly the same profile, and the shape of the overall spectrum
differs little from that obtained in traditional dark-matter models.
   
By contrast, the discrepancy between the recoil-energy spectra associated 
with DDM ensembles and those associated with traditional dark-matter models can
be quite striking for small $m_0$.  In particular, two distinctive 
features emerge which serve to 
distinguish the recoil-energy spectra associated with DDM ensembles from those 
associated with traditional dark-matter candidates in this regime.  The first of 
these is an apparent ``kink'' in the spectrum which arises for $m_0\lesssim 20$~GeV and 
large $\Delta m$.  Physically, this kink occurs because the 
contribution from $\chi_0$ to the 
differential event rate dominates at small $E_R$, but falls sharply as $E_R$ increases. 
By contrast, the contribution from each of the remaining, heavier $\chi_j$ falls far less 
sharply with recoil energy; hence these contributions collectively dominate at large $E_R$.      
The kink represents the transition point between these two $E_R$ regimes.  Similar
kinks also arise, for example, in the recoil-energy spectra of two-component dark-matter 
models~\cite{ProfumoTwoComponentDirectDet}.
  
The second distinctive feature which appears in the recoil-energy spectra displayed in
Figs.~\ref{fig:dRdERPanelsXe} and~\ref{fig:dRdERPanelsGe} 
emerges in cases in which $m_0$ and $\Delta m$ are both 
quite small.  In this case, a large number of the $\chi_j$ are sufficiently light that 
the profiles of their individual contributions to the differential event rate depend 
quite sensitively on $m_j$.  Moreover, these individual contributions cannot be 
resolved for small $\Delta m$; rather, they collectively conspire to produce an 
upturning (indeed, an upward concavity) of the recoil-energy spectrum at low $E_R$.  
The characteristic ``S''-shaped or ``ogee''-shaped curve which results from 
this upturning is most strikingly 
manifest in the $\Delta m = \{1,10\}$~GeV curves in the left panel of each figure.  
This ogee shape is a distinctive feature of DDM ensembles and is difficult to realize 
in traditional dark-matter models or in multi-component dark-matter 
models involving only a small number of dark-sector fields. 
As we shall see in Sect.~\ref{sec:Prospects}, both the kink and ogee features highlighted
here can serve to distinguish DDM ensembles at future direct-detection experiments.


\section{Constraining DDM Ensembles with Current Direct-Detection Data\label{sec:Limits}}


We begin our analysis of the direct-detection phenomenology of DDM ensembles
by assessing how current experimental data constrain the parameter space 
of our simplified DDM model.
The most stringent limit on spin-independent interactions between 
dark matter and atomic nuclei is currently that established by the 
XENON100 experiment~\cite{XENON100Paper2011} on the basis of 224.56 
live days of observation~\cite{XENON100Paper2012} 
with a fiducial mass of 34~kg of liquid xenon.
Two events were observed within the recoil-energy window
$6.6\mathrm{~keV} \leq E_R \leq 30.6\mathrm{~keV}$
which passed all cuts, and $1.0 \pm 0.2$ background events were expected.
Under the standard assumptions about the velocity distribution of the 
particles in the dark-matter halo outlined in Sect.~\ref{sec:ScatteringTradDM},
\etc, this result excludes at 90\%~C.L.\ {\it any}\/ dark-matter candidate ---
be it a traditional candidate or a DDM ensemble --- for which the 
total rate for nuclear recoils with $E_R$ within this recoil-energy window 
fails to satisfy the constraint   
\begin{equation}
  R ~\lesssim~ 4.91 \times 10^{-4}\mathrm{~kg}^{-1}\mathrm{~day}^{-1}~.
  \label{eq:GenericXENON100RateConstraint}
\end{equation}

In any arbitrary DDM model,
the total event rate $R$ for nuclear recoils observed at a given detector is 
obtained by integrating the differential rate in Eq.~(\ref{eq:DiffRateTotalGeneral})
over the range of $E_R$ values which fall within the particular energy window
$\Emin \leq E_R\leq \Emax$ established for that detector.
The contribution to this total rate from each $\chi_j$ is also scaled 
by an acceptance factor $\mathcal{A}_j(E_R)$ which depends both on its 
mass $m_j$ and on recoil energy.  In our simplified DDM model, we therefore have        
\begin{equation}
  R~=~
    \frac{\sigmaSI_{n0}\rho_{\mathrm{tot}}^{\mathrm{loc}}A^2}
    {2 \mu_{n0}^2 m_0}
     (1-\eta) \int_{\Emin}^{\Emax} dE_R 
     F^2(E_R)\,
     \sum_{j=0}^\infty \mathcal{A}_j(E_R)I_j(E_R)
     \left(1+j^\delta\frac{\Delta m}{m_0}\right)^{\alpha+2\beta-1}~.
  \label{eq:IntRateTotal}
\end{equation}   
In practice, the dependence of $\mathcal{A}_j(E_R)$ on both $E_R$ and $m_j$ 
tends to be slight over the range of $E_R$ values typically considered in 
noble-liquid and solid-state detectors.  We therefore approximate 
$\mathcal{A}_j(E_R) \approx 0.5$ for all $\chi_j$ in our analysis of the 
XENON100 constraint, in accord with the acceptance values quoted in
Ref.~\cite{XENON100Paper2011}.

\begin{figure}[h!]
\begin{center}
  \epsfxsize 2.25 truein \epsfbox
    {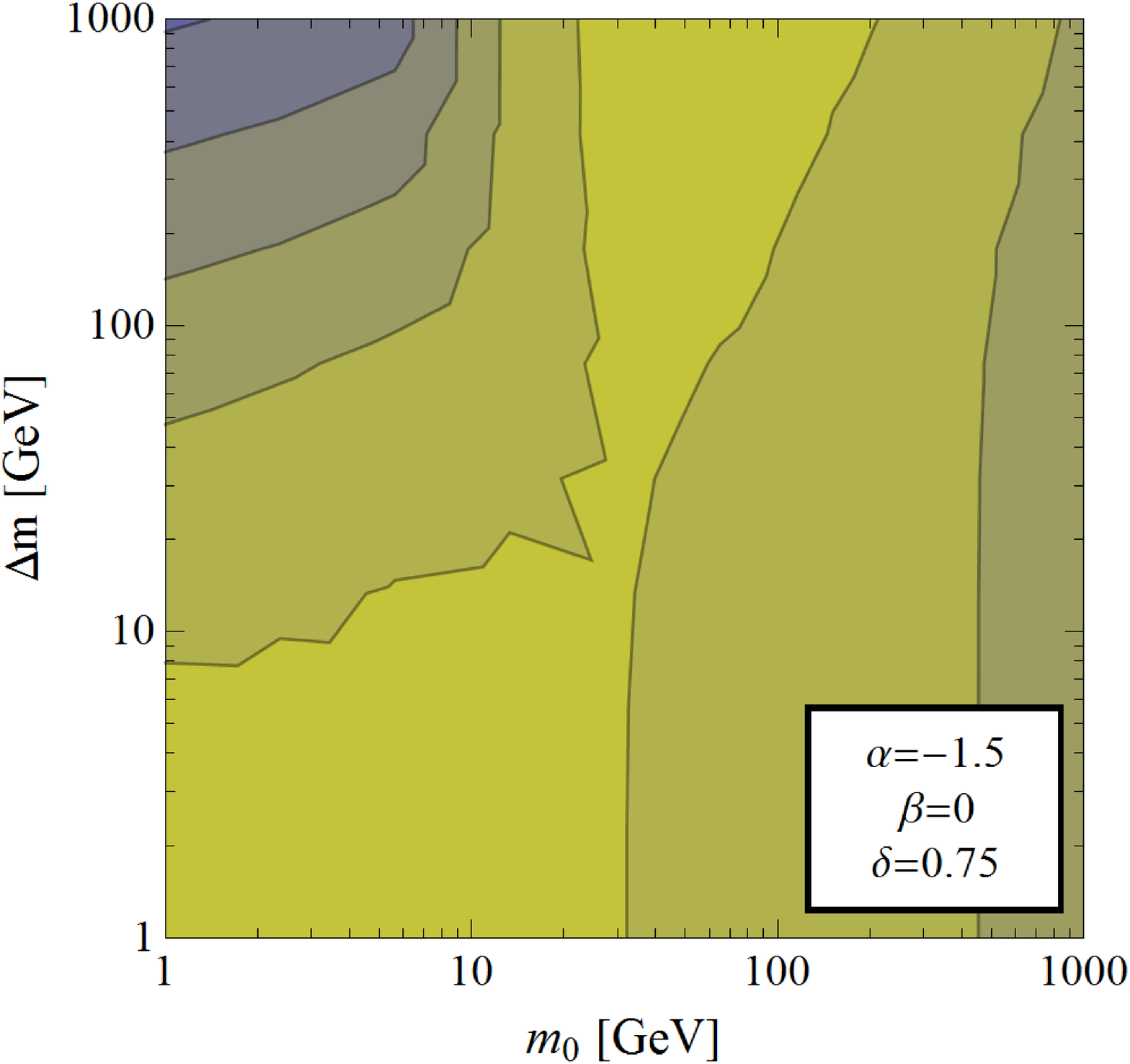}
  \epsfxsize 2.25 truein \epsfbox 
    {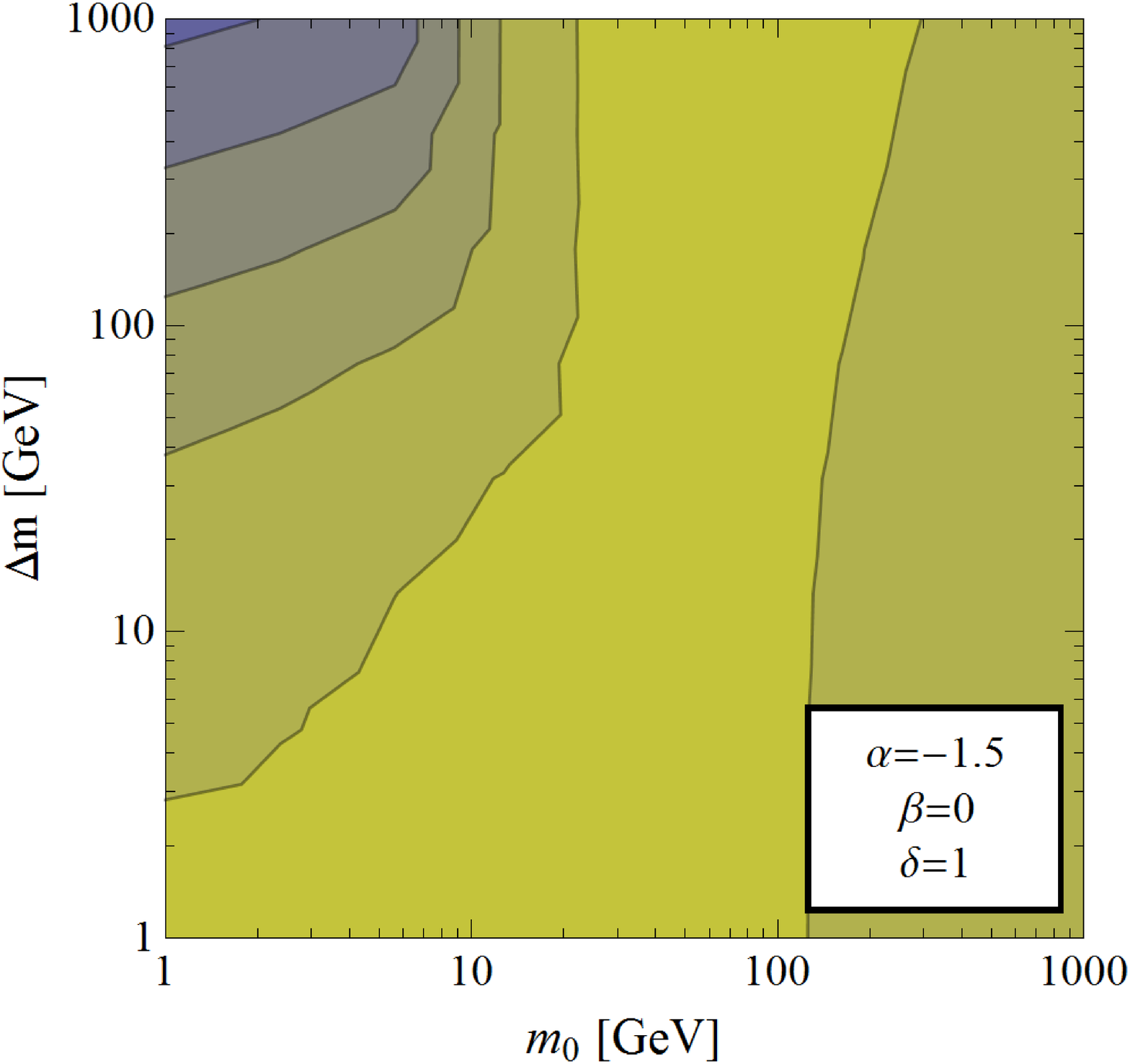}
  \epsfxsize 2.25 truein \epsfbox 
    {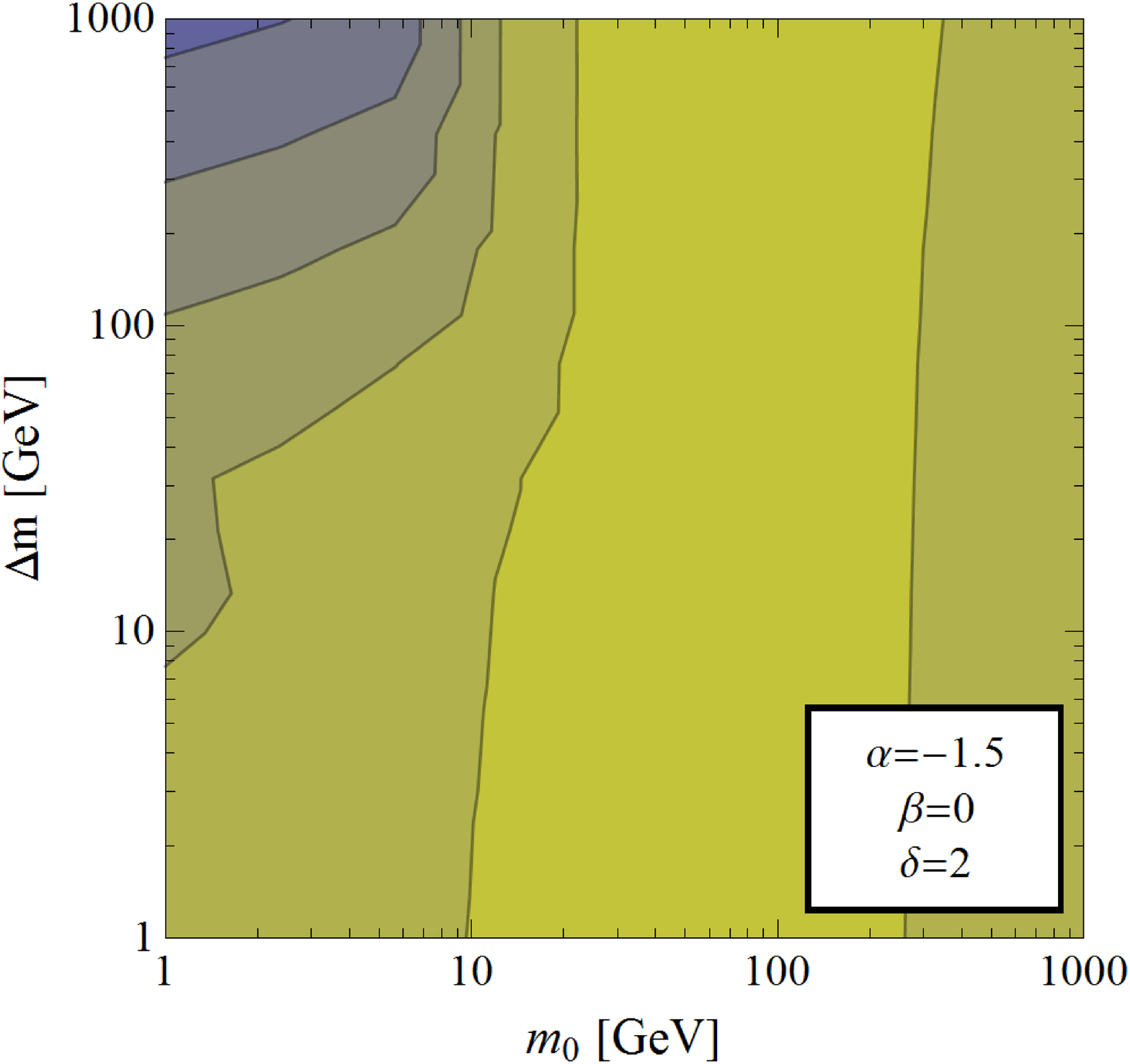} \\
  \epsfxsize 2.25 truein \epsfbox 
    {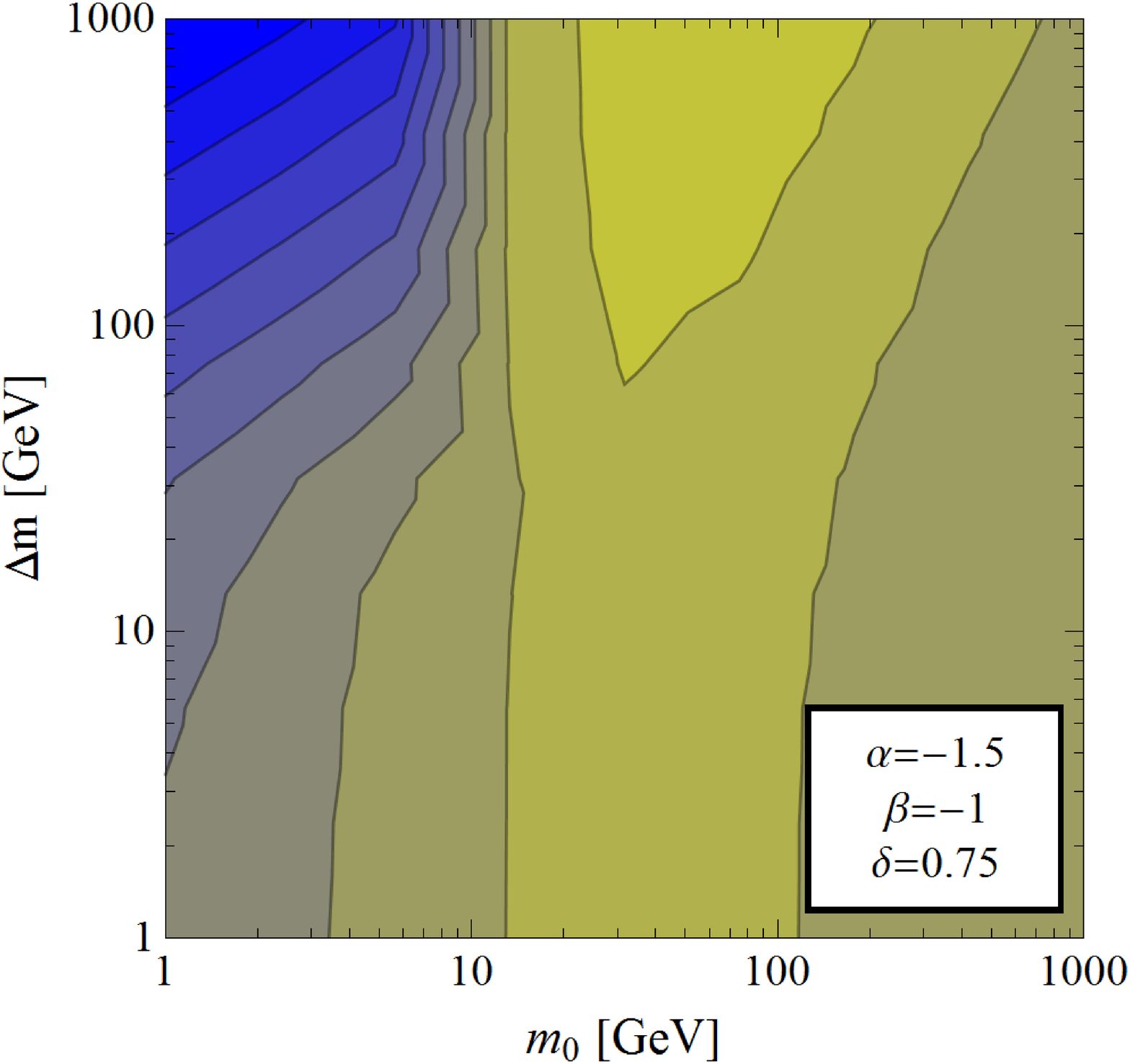}
  \epsfxsize 2.25 truein \epsfbox 
    {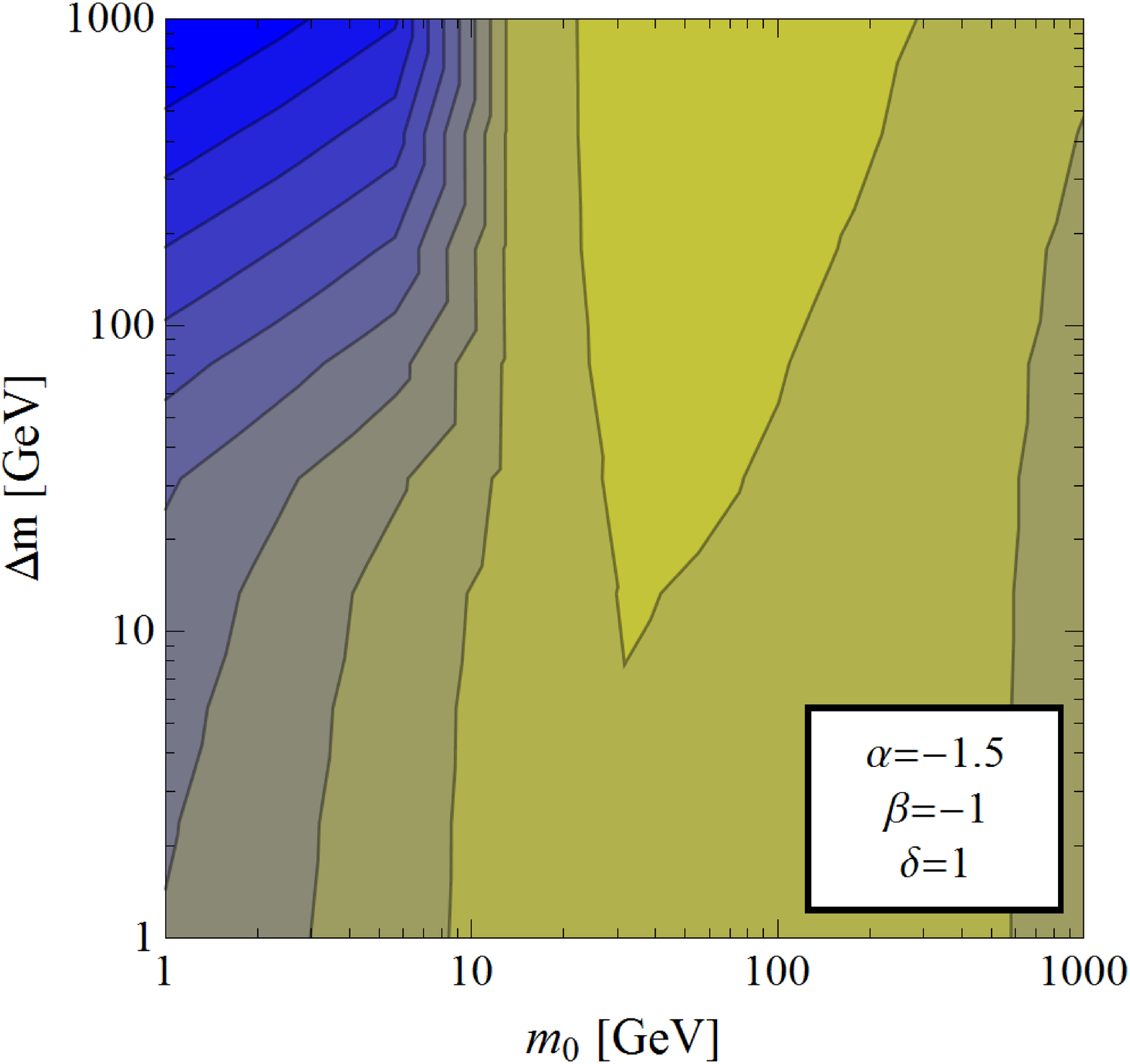}
  \epsfxsize 2.25 truein \epsfbox 
    {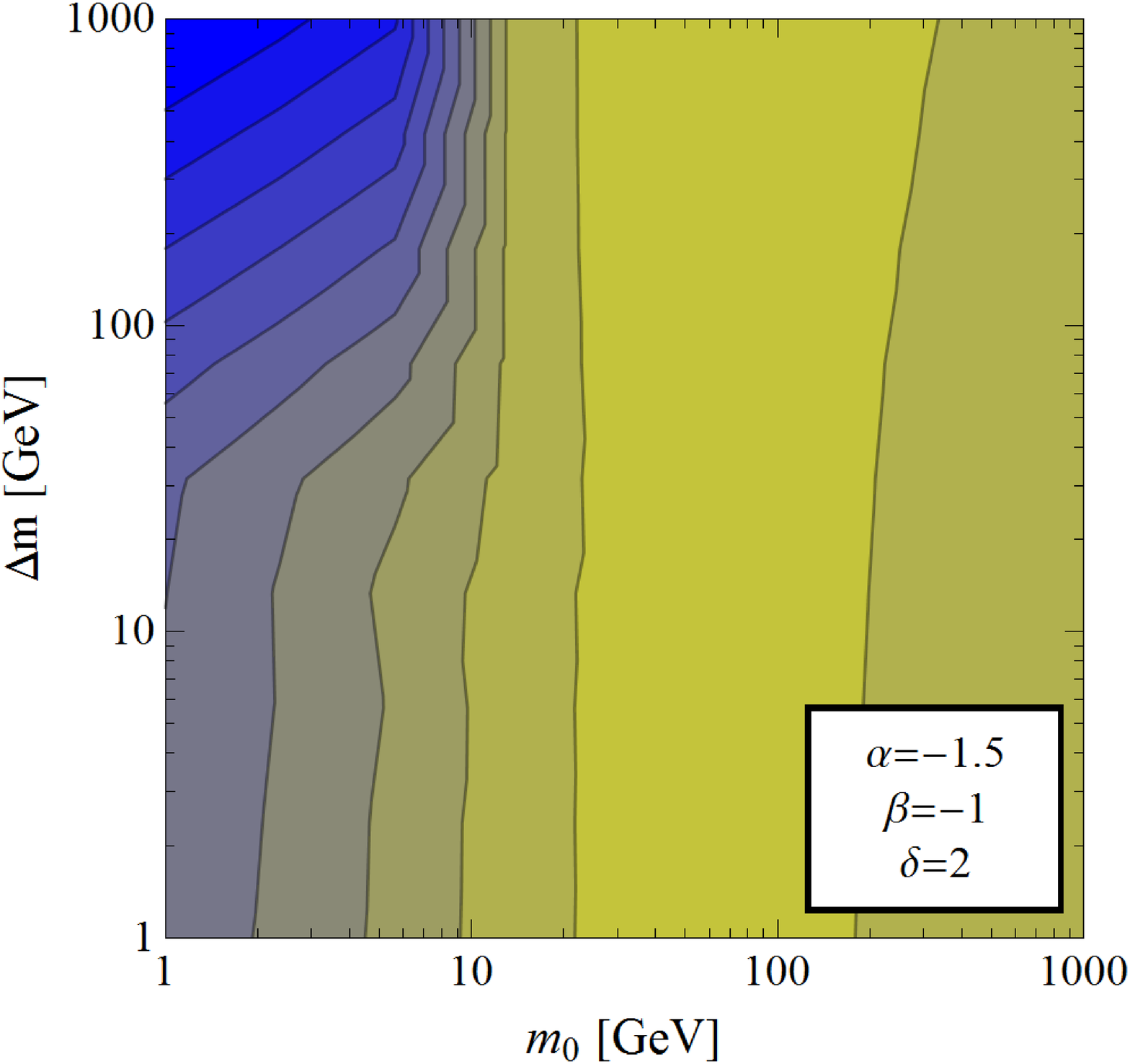} \\
  \epsfxsize 2.25 truein \epsfbox 
    {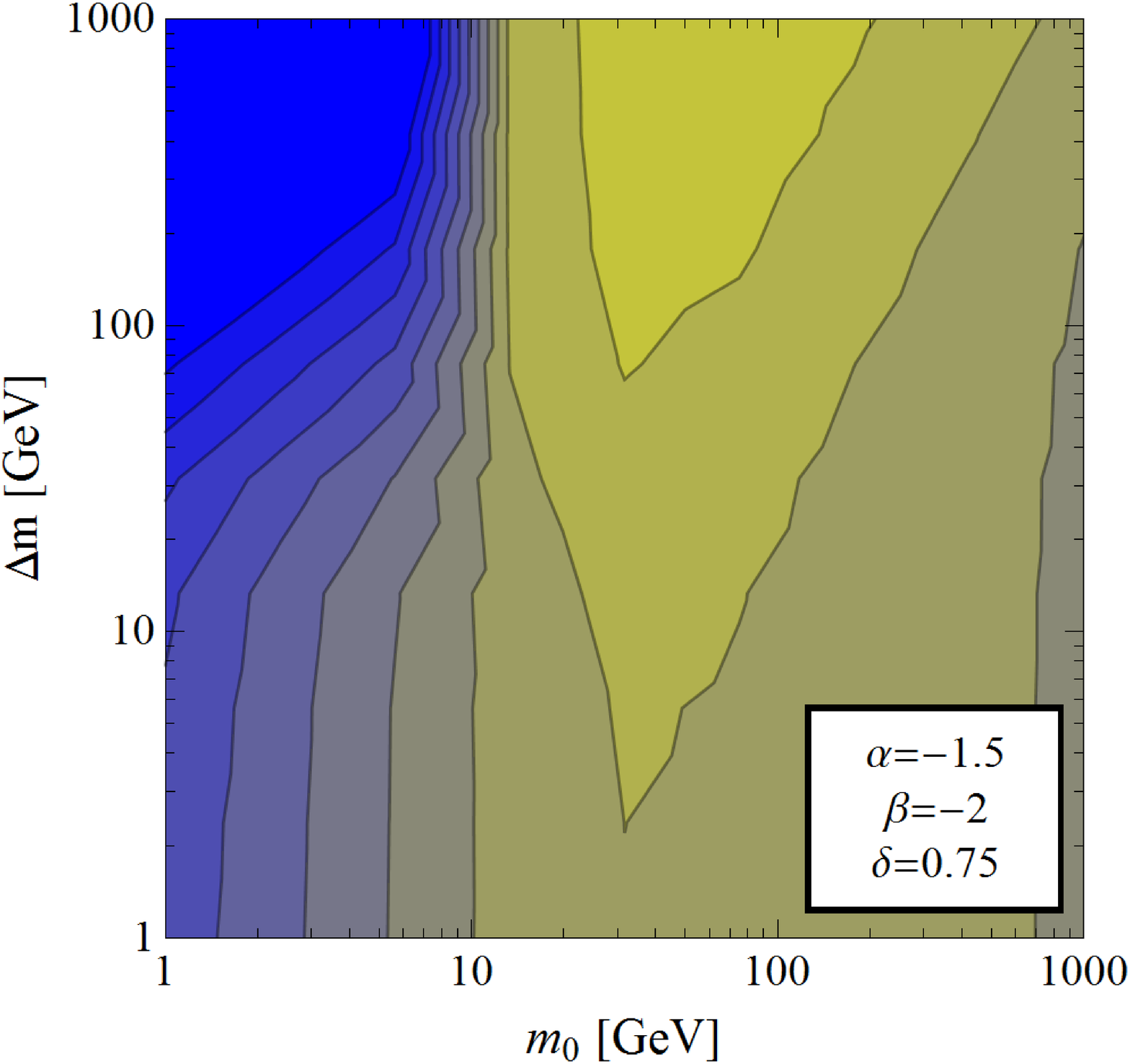}
  \epsfxsize 2.25 truein \epsfbox 
    {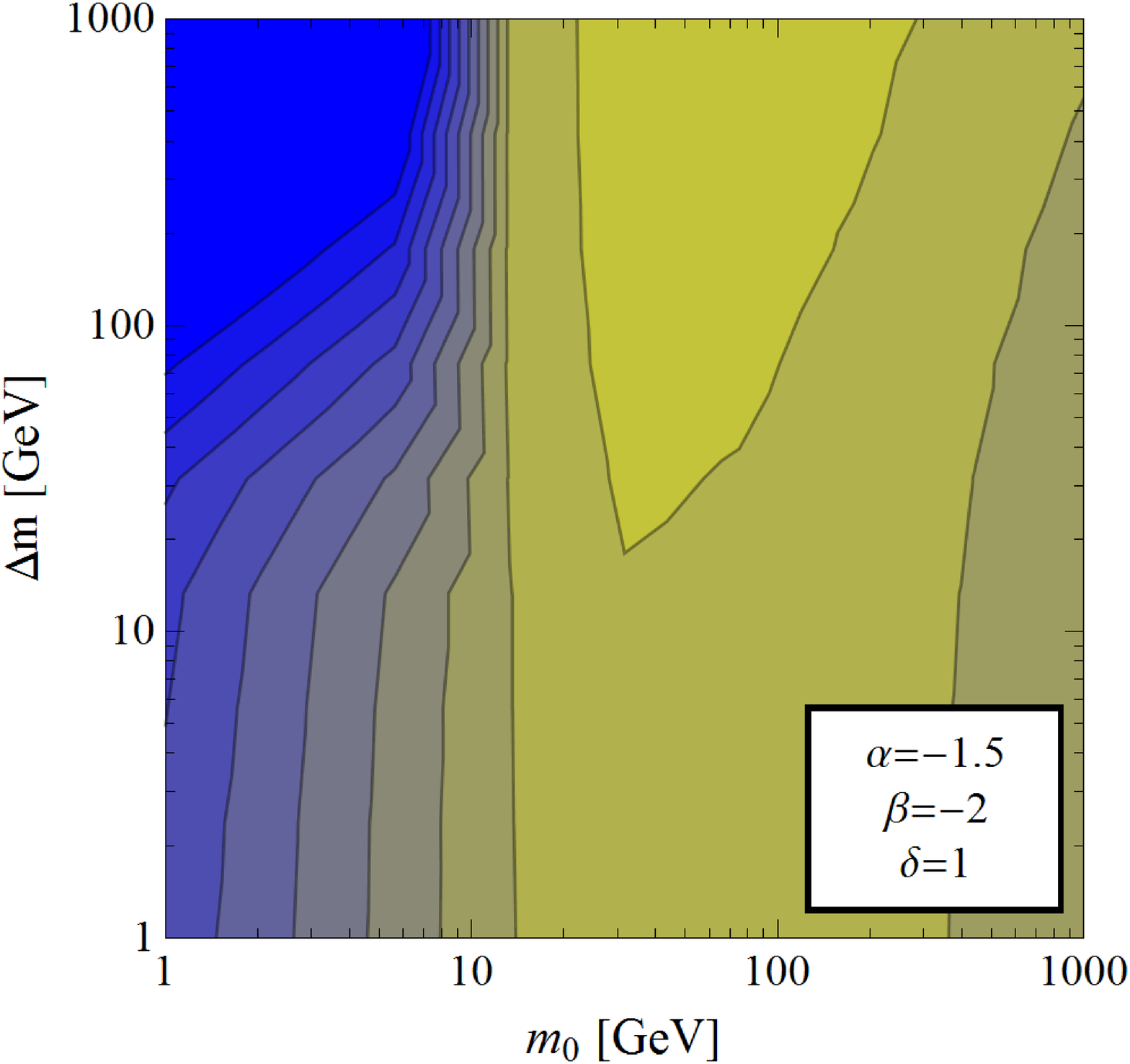}
  \epsfxsize 2.25 truein \epsfbox 
    {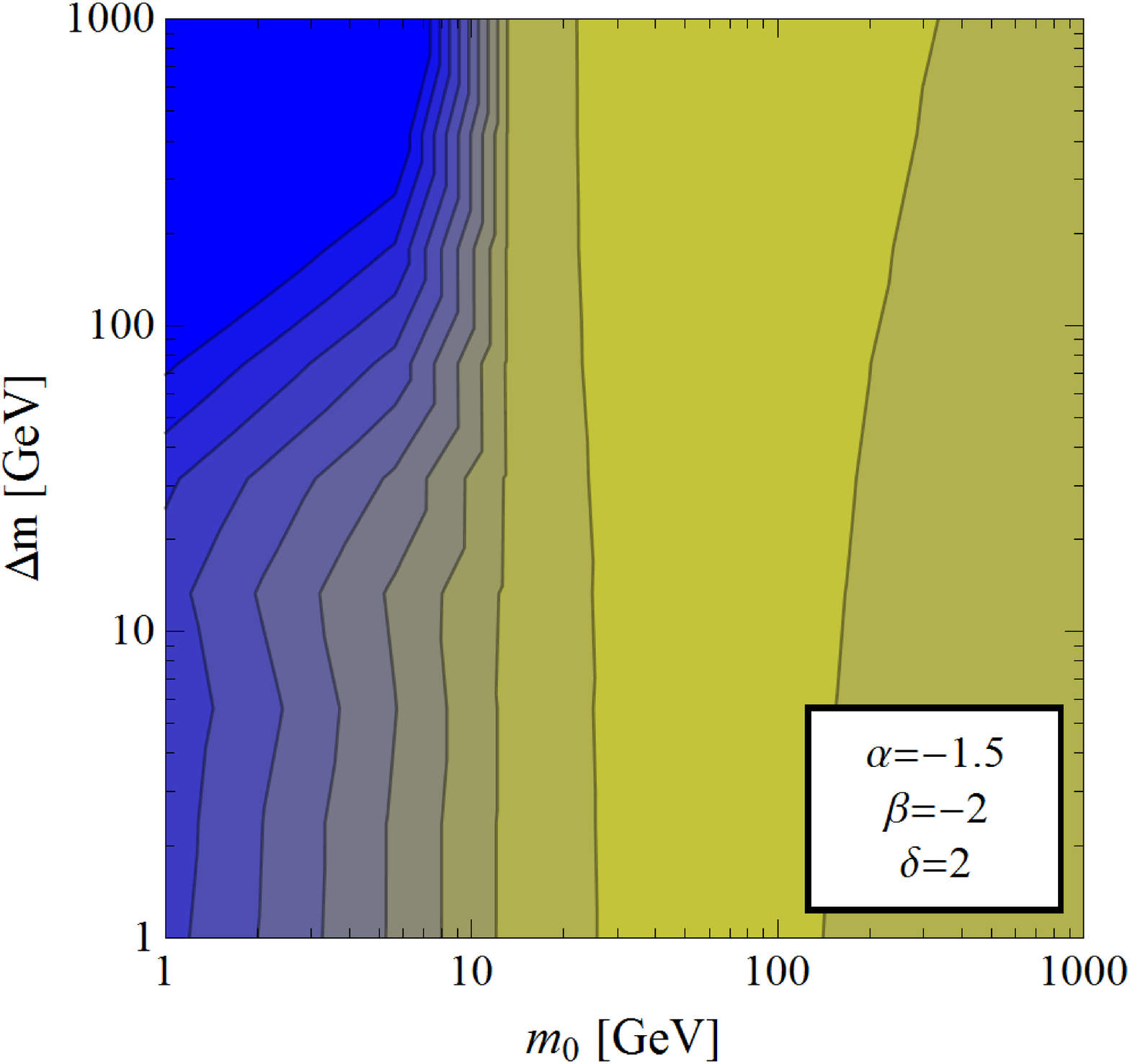} \\
  \raisebox{0.5cm}{\large $\sigma^{(\mathrm{SI})}_{n0,\mathrm{max}}~[\mathrm{cm}^{-2}]$~:~~}
     \epsfxsize 5.00 truein \epsfbox {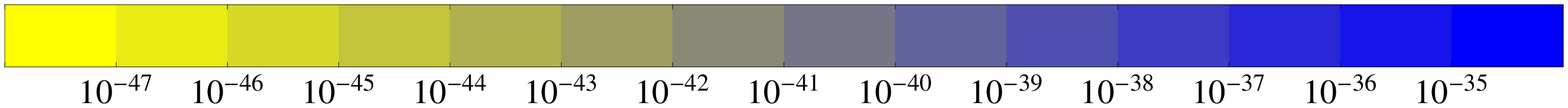}
\end{center}
\caption{
Contour plots showing the 90\%~C.L.\ limit from XENON100 on the spin-independent 
cross-section per nucleon $\sigmaSI_{n0}$ (in $\mathrm{cm}^{-2}$) of the 
lightest constituent particle $\chi_0$ in our simplified DDM model.  
The panels appearing in the top, center, and bottom rows show results for 
$\beta = \{0,-1,-2\}$, respectively.  The panels appearing in the 
left, center, and right columns show the results for $\delta = 0.75, 1, 2$,
respectively.  In each panel, we have set $\alpha = -1.5$.   
\label{fig:SigmaSILimitPanels}}
\end{figure}

In Fig.~\ref{fig:SigmaSILimitPanels}, we display a series of contour plots
showing the 90\%~C.L.\ limit in Eq.~(\ref{eq:GenericXENON100RateConstraint})
expressed as a bound on $\sigmaSI_{n0}$ in our simplified DDM model.  Of course,
for large $\Delta m$, the 90\%~C.L.\ limit value of $\sigmaSI_{n0}$
approaches the limit~\cite{XENON100Paper2012} on $\sigmaSI_{n\chi}$  
for a traditional dark-matter candidate with mass $m_\chi \approx m_0$.
However, when $\Delta m$ is small and a larger number of states 
contribute significantly to the total event rate, the experimental limit 
can differ substantially from that obtained for a traditional dark-matter 
candidate.  Such deviations become particularly pronounced for 
$m_0 \lesssim 10$~GeV, in which case the majority of nuclear-recoil events 
initiated by the lightest constituent $\chi_0$ in the ensemble have 
$E_R$ values which lie below the detector threshold.  In this region, 
the heavier $\chi_j$ collectively provide the dominant contribution to the
total event rate.  However, it is evident from 
Fig.~\ref{fig:SigmaSILimitPanels} that the heavier $\chi_j$ can also play an 
important role in the direct-detection phenomenology of DDM models even 
in the regime in which $m_0 \gtrsim 10$~GeV.


\section{Distinguishing DDM Ensembles at Future Detectors\label{sec:Prospects}}


We now examine the potential for distinguishing DDM ensembles from 
traditional dark-matter candidates at future direct-detection experiments.
Of course, an initial discovery of either a DDM ensemble or a traditional 
dark-matter candidate at a given direct-detection experiment would take 
the form of an excess in the total number of nuclear-recoil events observed 
above the expected background.  Our principal aim is therefore to determine 
the degree to which replacing the traditional dark-matter 
candidate with a DDM ensemble --- keeping all other aspects of our 
standard benchmark unchanged --- would result in a discernible deviation in the 
recoil-energy spectra measured at such experiments, once such an excess 
is observed.  For concreteness, we consider the situation in which the total 
scattering rate lies just below the sensitivity of current experiments, so 
that a sizable number of signal events is observed.             

Our procedure for comparing the recoil-energy spectrum associated with a 
given DDM ensemble to the spectrum associated with a traditional dark-matter 
candidate with mass $m_\chi$ is analogous to that used in
Ref.~\cite{DDMColliders} to compare invariant-mass distributions at the LHC
in the corresponding theories.  Similar procedures were also used in  
Ref.~\cite{ProfumoTwoComponentDirectDet}.
In particular, we partition each of the two spectra into $n_b$ bins with 
widths greater than or equal to the recoil-energy resolution $\Delta E_R$ 
at the minimum $E_R$ in the bin.  We then construct the $\chi^2$ statistic 
\begin{equation}
  \chi^2(m_{\chi}) ~=~ \sum_k\frac{[X_k - \mathcal{E}_k(m_\chi)]^2}{\sigma_k^2}~,
\end{equation}
where the index $k$ labels the bin, $X_k$ is the expected population of events in 
bin $k$ in the DDM model, $\mathcal{E}_k(m_\chi)$ is the expected population of events in
bin $k$ in the traditional dark-matter model, and $\sigma_k^2$ is the variance in 
$X_k$ due to statistical uncertainties.
Since the $X_k$ are distributed according to a multinomial 
distribution, it follows that $\sigma_k^2 = X_k (1-X_k/N_e)$, where $N_e$ denotes 
the total number of signal events observed.

The proper measure of the distinctiveness of the recoil-energy spectrum 
associated with a DDM ensemble is not the degree to which it differs from that 
associated with a traditional dark-matter candidate with a particular $m_\chi$, 
but rather from {\it any} such dark-matter candidate.
Consequently, we survey over traditional dark-matter candidates $\chi$ 
with different values of $m_\chi$ with all other assumptions
held fixed.  Note that the total event rate $R$ --- and hence also 
the spin-independent cross-section per nucleon 
$\sigmaSI_{n\chi}$ for each value $m_\chi$ --- is effectively specified 
by the signal-event count $N_e$; thus $m_\chi$ is the only remaining
parameter over which we must survey.  We then take
\begin{equation}
  \chi^2_{\mathrm{min}} ~\equiv~ \min_{m_\chi} \big\{\chi^2(m_\chi)\big\}
  \label{eq:ChiSqMinDef}
\end{equation}
as our measure of the distinctiveness of the 
recoil-energy spectrum associated with a given DDM ensemble.  
We then evaluate a statistical significance of differentiation in each case 
by comparing $\chi^2_{\mathrm{min}}$ to a $\chi^2$ distribution with $n_b-1$
degrees of freedom.  Specifically, this is defined to be the significance to 
which the $p$-value obtained from this comparison would correspond for a Gaussian 
distribution.

For this study, we choose not to limit our attention to any particular experiment, 
either existing or proposed; rather, we investigate the 
prospects for distinguishing DDM ensembles at a pair of hypothetical
detectors, each with characteristics representative of a particular class of 
next-generation direct-detection experiments.  The first of these is a dual-phase 
liquid-xenon detector with attributes similar to those projected for XENON1T 
and future phases of the LUX experiment.  The other is a germanium-crystal 
detector with attributes similar to those projected for GEODM and 
the SNOLAB phase of the SuperCDMS experiment.
For both experiments, we assume five live years of data collection time.

Each of these hypothetical detectors is characterized by its recoil-energy 
resolution $\Delta E_R$, recoil-energy window, signal acceptance, fiducial
mass, and the differential event rate associated with the combined background.  
For our hypothetical next-generation xenon detector, we choose a recoil-energy 
window $8\mathrm{~keV}\leq E_R\leq 48\mathrm{~keV}$, a fiducial mass of 5000~kg, 
and a signal acceptance $\mathcal{A}_j(E_R) \approx 0.5$ which is independent of 
both $E_R$ and $m_j$.  We model the energy resolution $\Delta E_R$ of our detector 
after that obtained for the combined S1 and S2 signals at the XENON100 experiment.    
This energy resolution is determined from measurements of the detector response
for a number of $\gamma$-ray calibration lines at various energies.  
The result, expressed in terms of electron-recoil-equivalent 
energy units $\keVee$, is~\cite{XENON100ExperimentDetails}
\begin{equation}
  \Delta E_R ~ \approx ~ 0.60 \times 
     \left(\frac{E_R}{\keVee}\right)^{1/2} \keVee~.
  \label{eq:EnergyResolutionXENON100keVee} 
\end{equation}
The corresponding energy resolution for nuclear recoils is related to this 
result by an energy-dependent effective quenching factor 
\begin{equation}  
 Q_{\mathrm{eff}}(E_R) ~ \equiv ~ 
   \frac{1}{\Leff(E_R)}\left(\frac{S_{\mathrm{ee}}}{S_{\mathrm{nr}}}\right),
\end{equation}
where $\Leff \equiv L_{\mathrm{nr}}^{(0)}/L_\mathrm{ee}^{(0)}$ is the ratio
of the light yield for nuclear recoils to that for electron recoils at zero 
applied electric field, and where $S_{\mathrm{ee}} = 0.58$ and $S_{\mathrm{nr}} = 0.95$ 
are electric-field-scintillation quenching factors which account for the effect 
of the $530\mathrm{~V/cm}$ applied drift field~\cite{XENON100SeeAndSnr}.  
The energy resolution $\Delta E_R$ for nuclear recoils
therefore depends on the uncertainties in $\Leff$, $S_{\mathrm{ee}}$, and $S_{\mathrm{nr}}$.   
For $E_R \gtrsim 3$~keV, the uncertainty in $\Leff$ is approximately independent of $E_R$ 
and given by $\Delta\mathcal{L}_{\mathrm{eff}} \approx 0.01$,
while the uncertainty in $S_{\mathrm{ee}}$ and $S_{\mathrm{nr}}$ is negligible.   
We therefore find that over the full recoil-energy window of our hypothetical detector, 
$\Delta E_R$ is given by       
\begin{equation}
  \Delta E_R ~ \approx ~ \left[0.36 \left(\frac{E_R}{\mathrm{keV}}\right)
      + \left(\frac{0.01}{\Leff(E_R)}\right)^2
      \left(\frac{E_R}{\mathrm{keV}}\right)^2 \right]^{1/2} \mathrm{keV}~,
  \label{eq:EnergyResolutionXENON100keVnr}
\end{equation}
where all energies are expressed in nuclear-recoil-equivalent units. 
In our analysis, the width of each bin is set equal to the value of 
$\Delta E_R$ at the lowest energy in the bin for this detector.

We model the differential event rate for the combined background at our 
hypothetical xenon detector after that projected for the combined 
background at XENON1T.  This background rate, after the application
of event-selection criteria (including both a multiple-scatter veto and
an S2/S1 cut), is dominated by electron-recoil events, and in particular
those from $^{85}\mathrm{Kr}$ and other impurities within the detector 
volume.  The recoil-energy spectrum associated with this background is 
approximately independent of $E_R$, and for a $^{85}\mathrm{Kr}$ 
concentration of 0.5~ppt is given by~\cite{XENON1TBGTalk}        
\begin{equation}
  \left(\frac{dR}{dE_R}\right)_{\mathrm{BG}} ~ \approx ~ 
    7\times 10^{-9}\mathrm{~kg}^{-1}\mathrm{day}^{-1}\mathrm{keV}^{-1}~.
\end{equation}  

For our hypothetical next-generation germanium detector, we likewise choose
the fiducial mass to be 5000~kg.  Moreover, we choose 
a recoil-energy acceptance window, energy resolution, and signal acceptance 
comparable with that of the CDMS~II experiment.  The recoil-energy 
acceptance window for CDMS~II is $10\mathrm{~keV} \leq E_R \leq 100\mathrm{~keV}$,  
and the energy resolution $\Delta E_R$ within this range is given
by~\cite{EnergyResolutionCompendium} 
\begin{equation}
  \Delta E_R ~\approx~ 0.2 \times \left(\frac{E_R}{\mathrm{keV}}\right)^{1/2}
  \mathrm{keV}~.
\end{equation}  
We adopt this same acceptance window and energy resolution for our hypothetical 
next-generation detector.  In order to avoid issues related to low statistics for 
this detector, we adopt a binning scheme coarser than its energy resolution would 
in principal allow.  In particular, we set the width of each bin equal to the 
$\Delta E_R$ value of our hypothetical xenon-based detector at the smallest $E_R$ 
value in the bin.  The signal acceptance for CDMS~II varies only slightly, from a 
minimum of $\mathcal{A}\approx 0.25$ at $E_R$ values near the endpoints $\Emin = 10$~keV 
and $\Emax = 100$~keV of the recoil-energy acceptance window to a maximum of
$\mathcal{A}\approx 0.32$ at $E_R \sim 20$~keV~\cite{CDMSIIEarlyData,CDMSIIFinalData}.
We therefore once again approximate the acceptance as independent of $E_R$ and $m_j$, 
and take $\mathcal{A}_j = 0.3$ for all $\chi_j$ for our hypothetical detector.   

In order to obtain a realistic recoil-energy spectrum for the combined background at 
our hypothetical germanium detector, we adopt the following procedure. 
We model the shape of this spectrum after that observed for the CDMS~II
experiment, which is dominated at $E_R \gtrsim 10$~keV by the contribution from 
surface events, and adopt a normalization such that the total event rate is  
$R_{\mathrm{BG}} \approx 1.0 \times 10^{-5}\mathrm{~kg}^{-1}\mathrm{~day}^{-1}$.  This
is a rate comparable to the background-event rate estimates for the SuperCDMS
detector at SNOLAB.  At recoil energies $E_R \gtrsim 10$~keV, 
the background rate at the CDMS~II detector is dominated by the contribution 
from surface events.  For $10\mathrm{~keV}\lesssim E_R \lesssim 25\mathrm{~keV}$, 
the recoil-energy spectrum is well modeled by~\cite{CDMSIIBackgrounds}  
\begin{equation}
  \left(\frac{dR}{dE_R}\right) ~\approx~ \left( 8.3 \times 10^{-7}\right)
  \times e^{-0.05\times (E_R/\mathrm{keV})}
  \mathrm{~kg}^{-1}\mathrm{day}^{-1}\mathrm{keV}^{-1}~.      
\end{equation}
From this result, we extrapolate the background spectrum over the full 
recoil-energy window of our detector.

We consider the situation in which the total rate for nuclear-recoil events 
engendered by the DDM ensemble lies just below the sensitivity of current 
experiments, in which case the number of signal events observed at the next 
generation of detectors will be substantial.  We examine the DDM differentiation
prospects at each of our hypothetical detectors independently, in isolation, 
rather than attempting to correlate the results between the two. 
For concreteness, we adopt a benchmark value of $N_e = 1000$ total 
signal events at each detector.  Note that for our chosen running time of 
five live years, this value of $N_e$ is consistent with the XENON100 limits 
discussed in Sect.~\ref{sec:Limits} throughout the entirety of the 
parameter space of our simplified DDM model which we include in our analysis
for both of our detectors.  

\begin{figure}[ht!]
\begin{center}
  \epsfxsize 2.25 truein \epsfbox
    {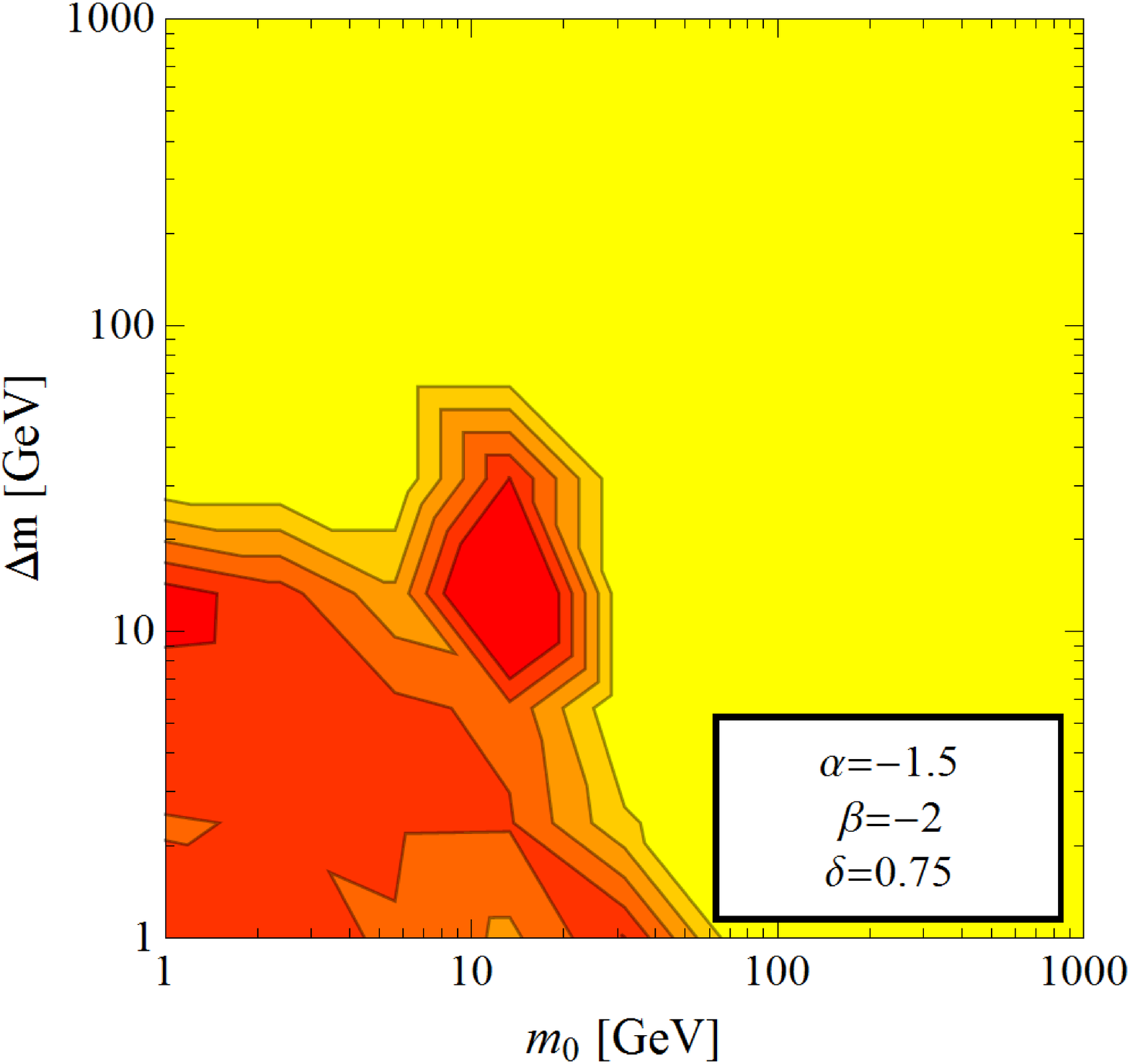}
  \epsfxsize 2.25 truein \epsfbox 
    {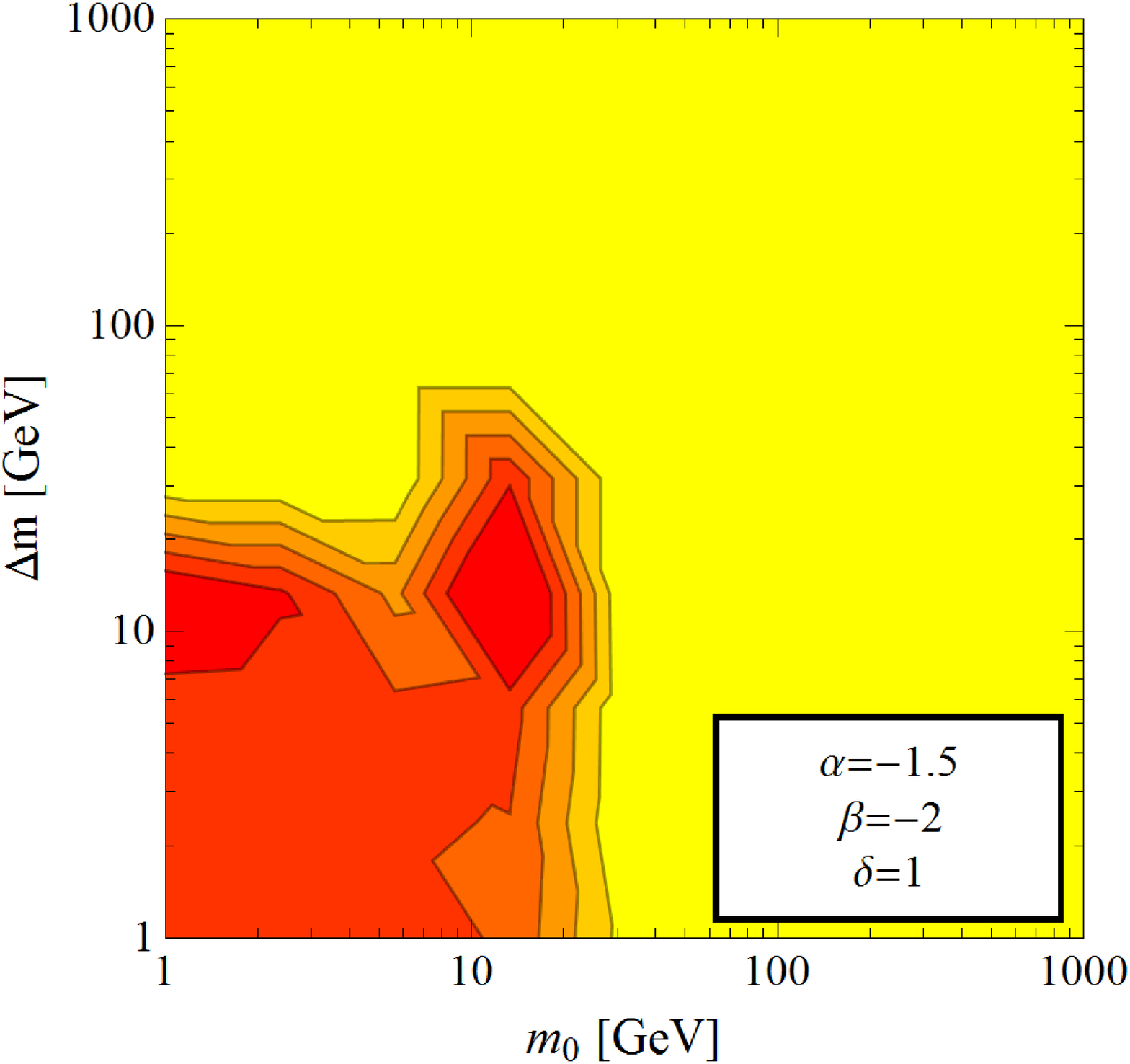}
  \epsfxsize 2.25 truein \epsfbox 
    {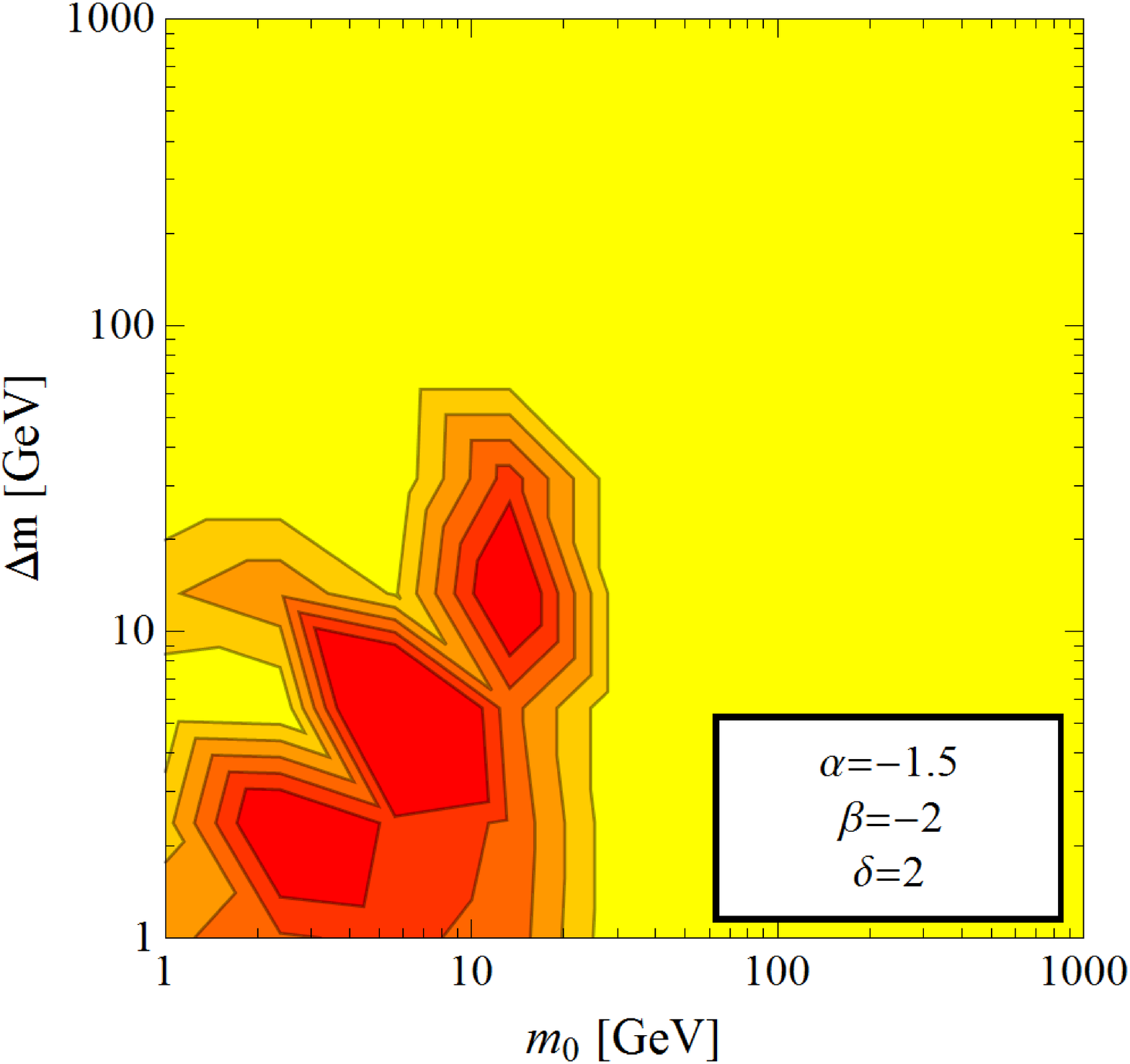}\\
  \epsfxsize 2.25 truein \epsfbox 
    {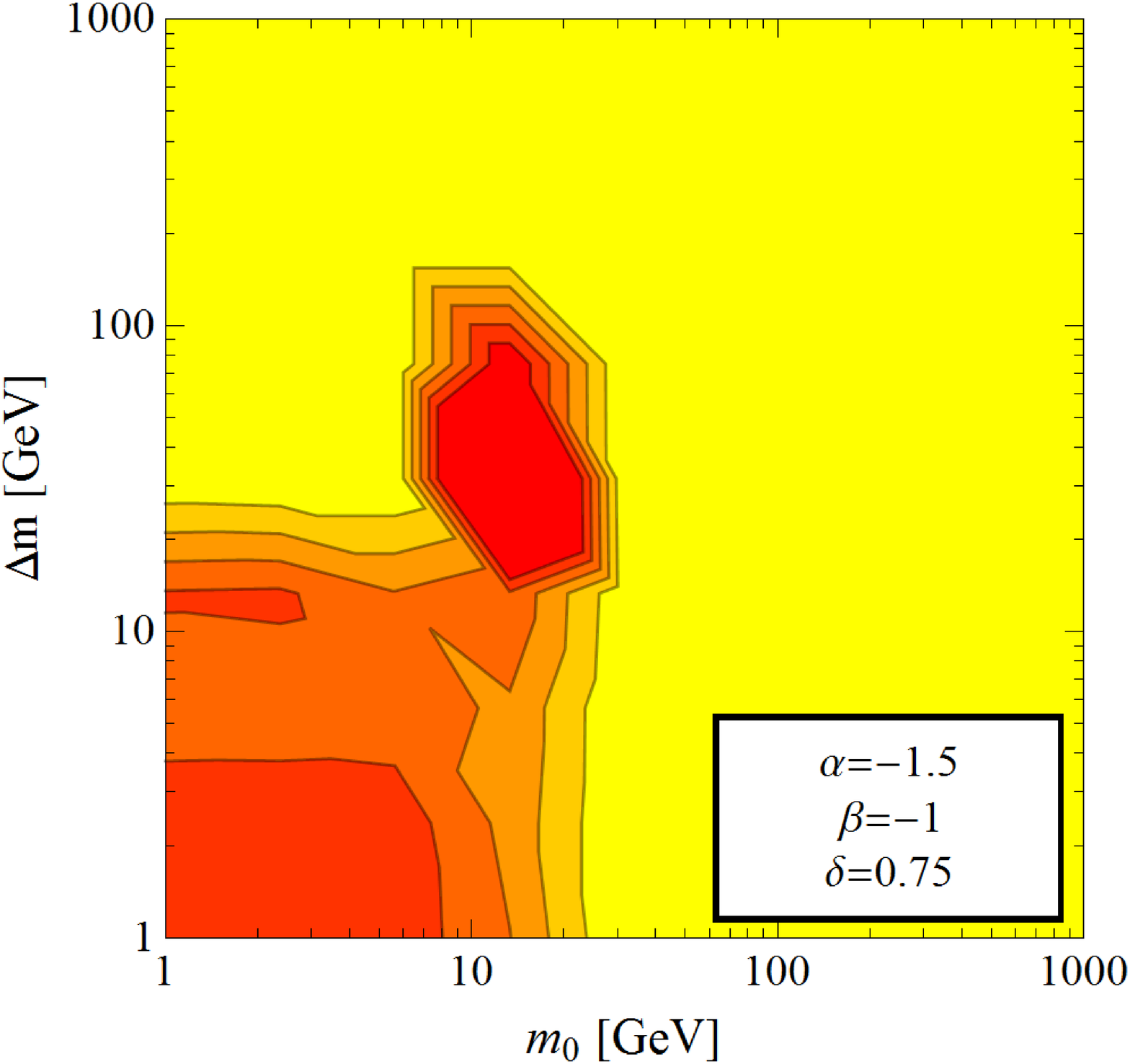}
  \epsfxsize 2.25 truein \epsfbox 
    {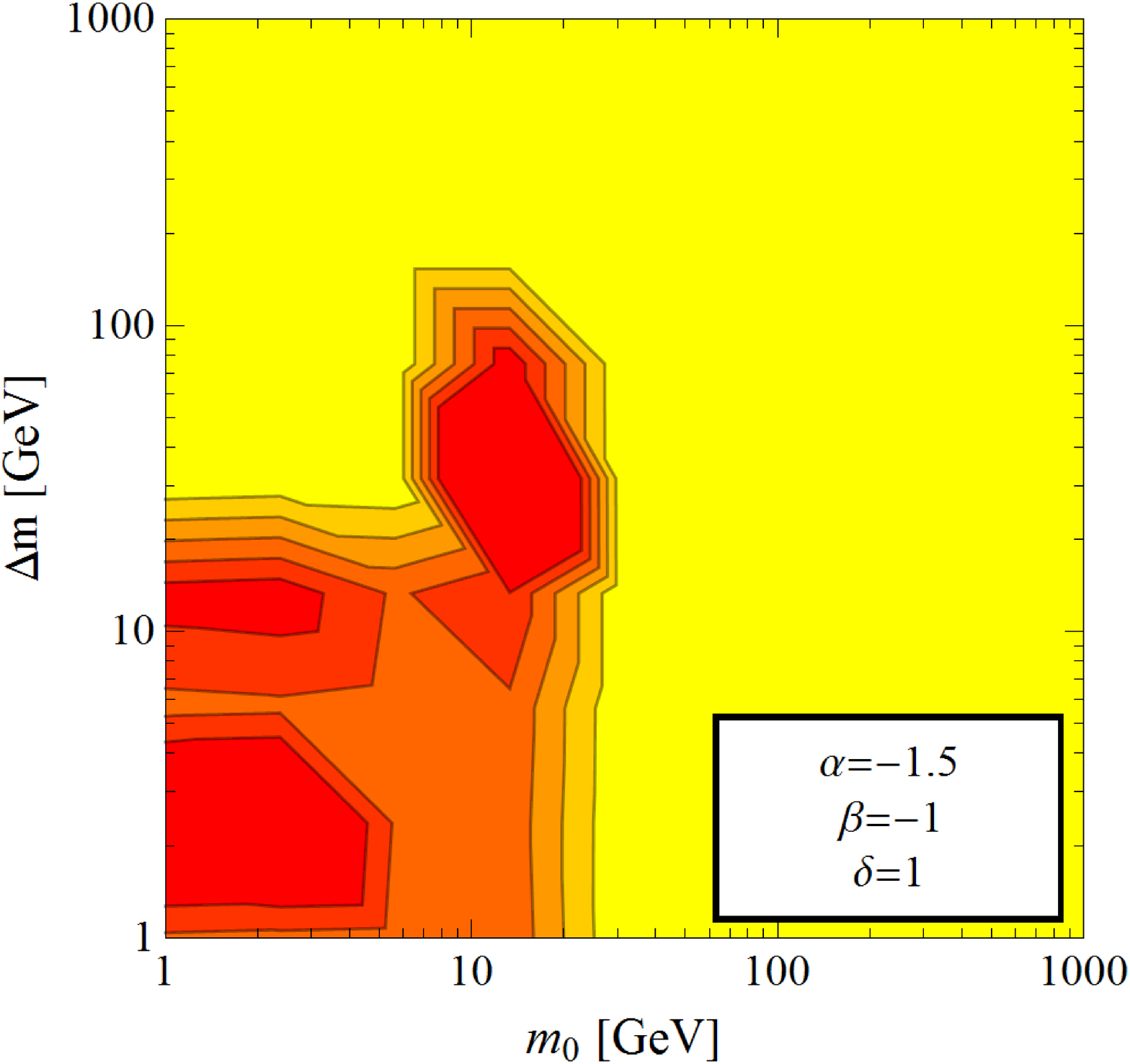}
  \epsfxsize 2.25 truein \epsfbox 
    {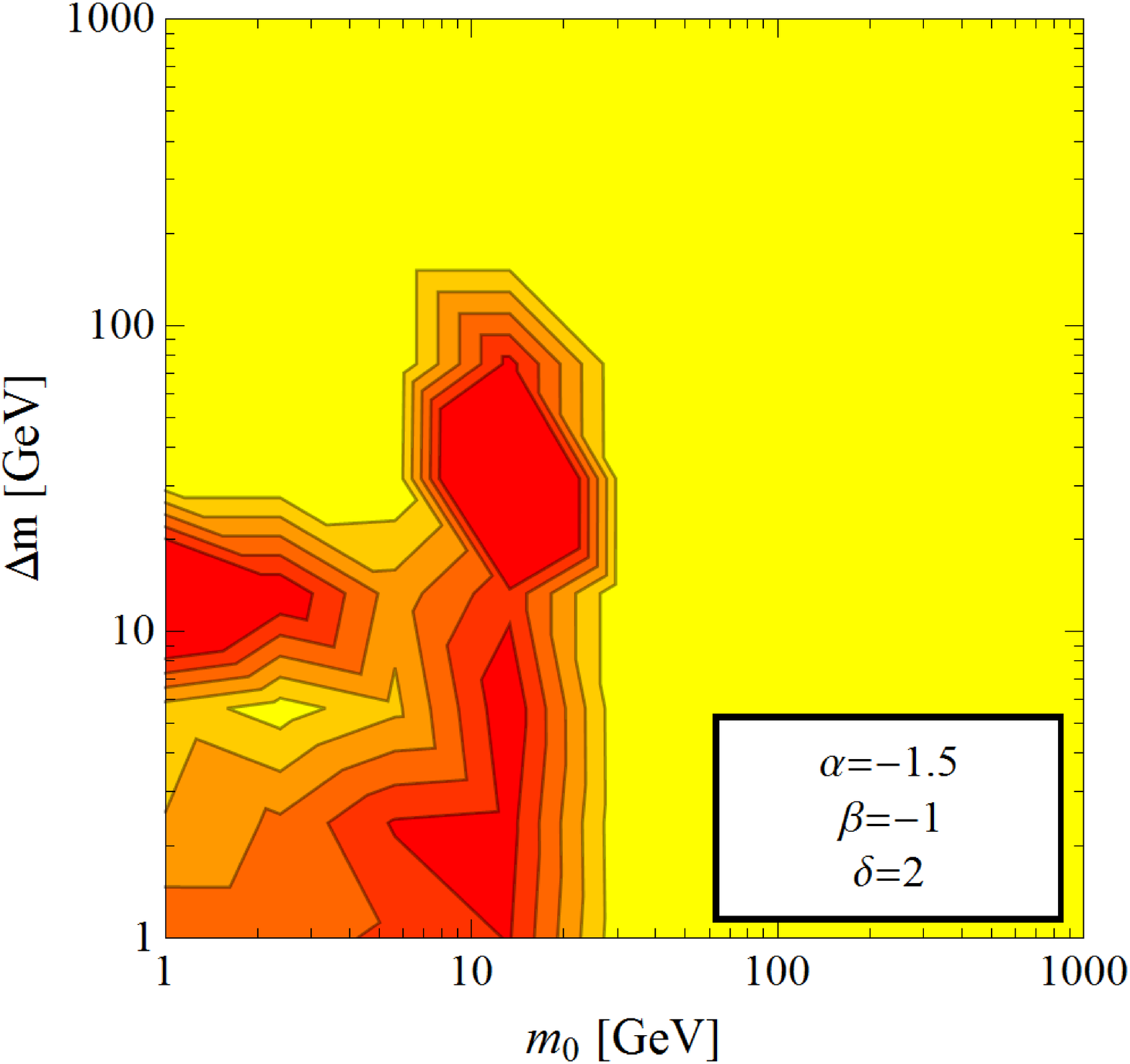}\\
  \epsfxsize 2.25 truein \epsfbox 
    {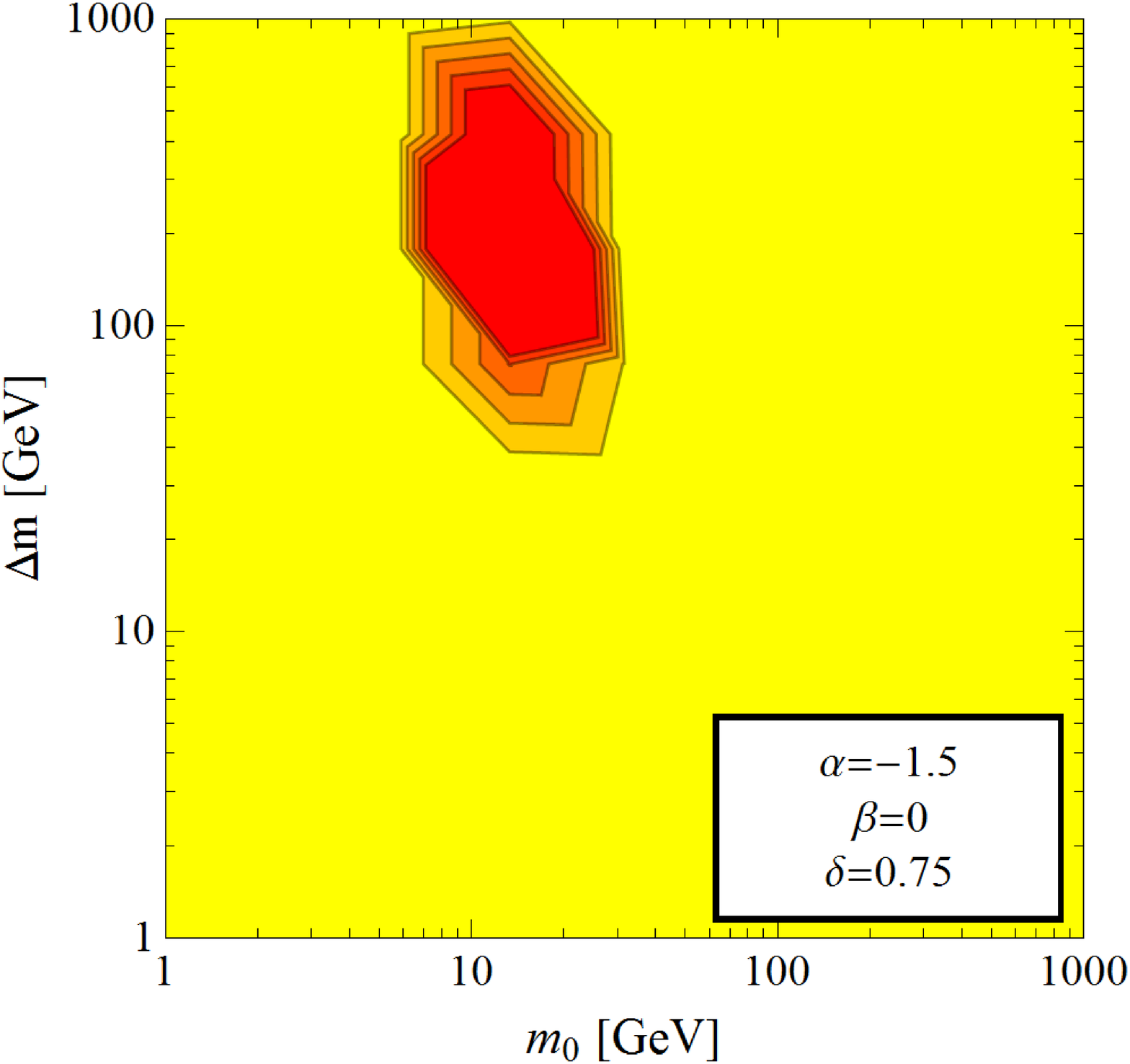}
  \epsfxsize 2.25 truein \epsfbox 
    {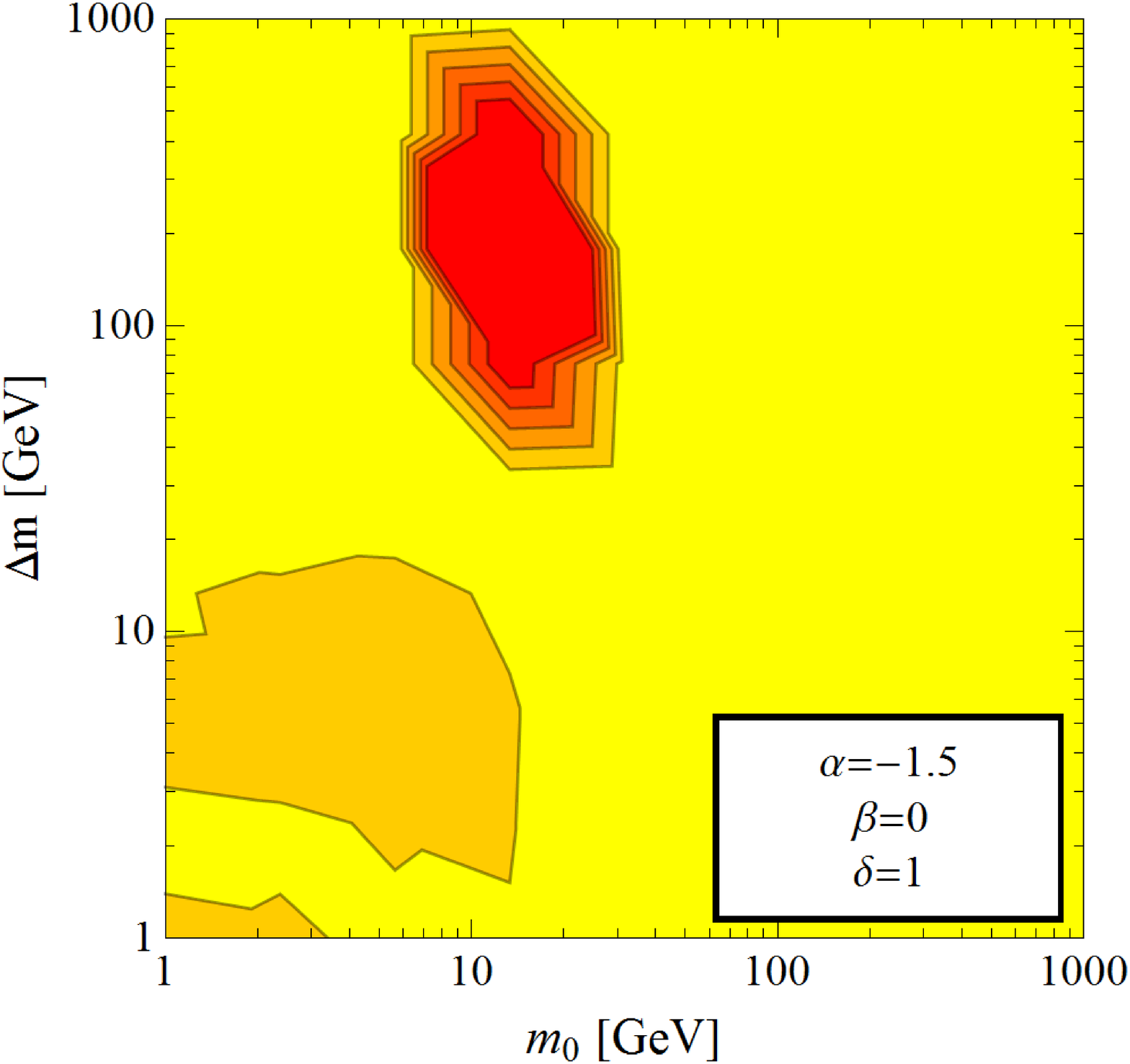}
  \epsfxsize 2.25 truein \epsfbox 
    {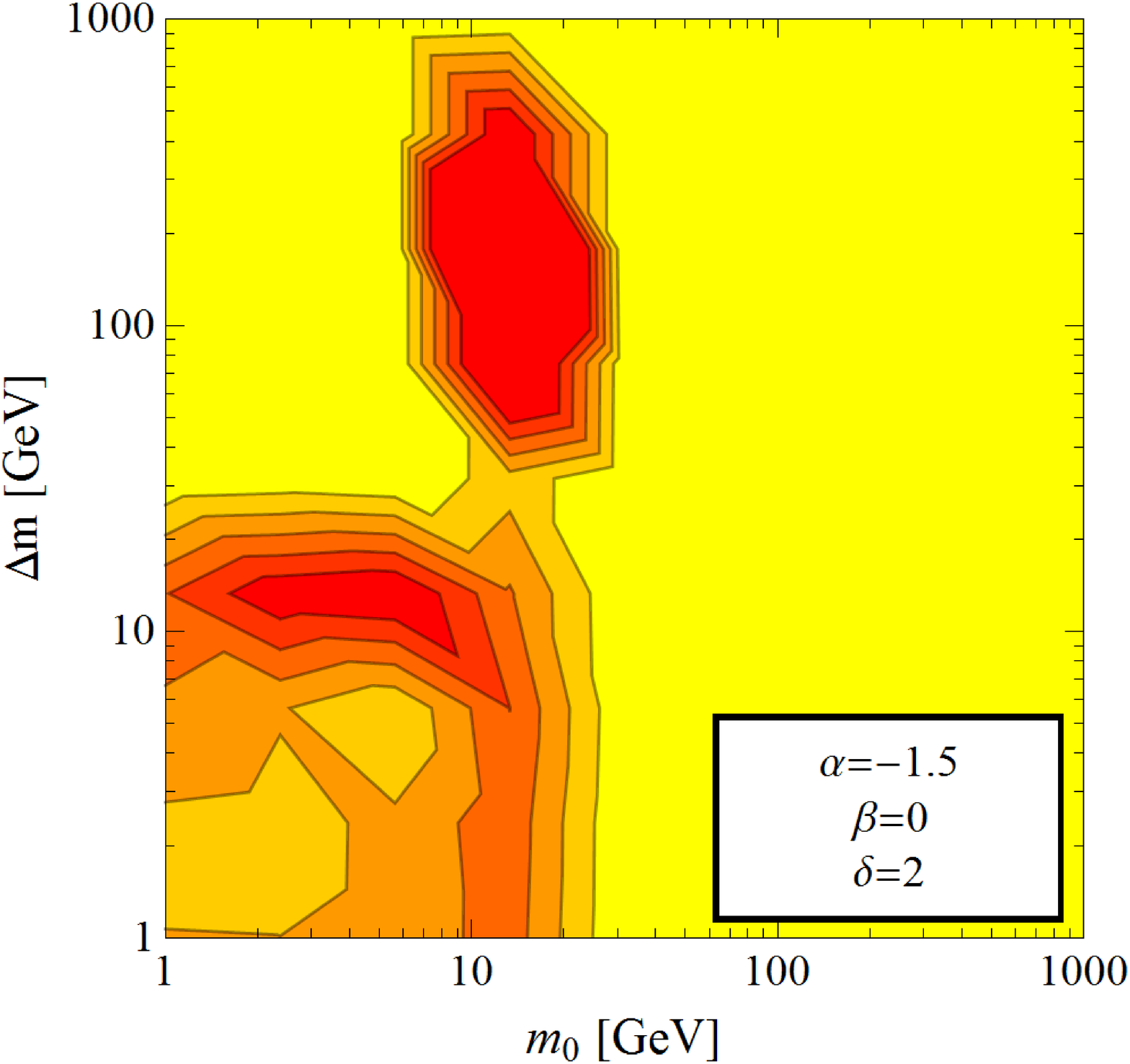}\\
  \raisebox{0.5cm}{\large Significance~:~~}
     \epsfxsize 5.00 truein \epsfbox {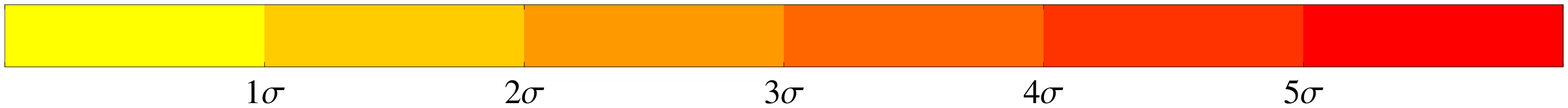}
\end{center}
\caption{
Contour plots showing the significance level at which the recoil-energy
spectrum associated with a DDM ensemble can be distinguished from that 
associated with {\it any}\/ traditional dark-matter candidate
which gives rise to the same total event rate at a hypothetical 
direct-detection experiment.  This experiment is taken to be a liquid-xenon 
detector with a fiducial volume of 5000~kg and characteristics otherwise similar 
to those of the proposed XENON1T experiment, as discussed in the text.  A 
running time of five live years and an event count of $N_e=1000$ signal events 
is assumed.        
\label{fig:SignificancePanelsXe}}
\end{figure}

In Fig.~\ref{fig:SignificancePanelsXe} we show how the projected statistical 
significance of differentiation obtained with $N_e = 1000$ signal events at 
our hypothetical xenon detector varies as a function of the parameters which 
characterize our simplified DDM model.  We find that from among these 
parameters, the significance is particularly sensitive to the values of
$m_0$ and $\Delta m$; hence we display our results in $(m_0,\Delta m)$ space, 
with $\alpha$, $\beta$, and $\delta$ held fixed in each of the panels shown.
The panels in the left, center, and right columns of the figure correspond 
respectively to $\delta = \{0.75,1,2\}$, and the panels in the top, center, 
and bottom rows correspond to $\beta = \{-2,-1,0\}$.  In all of the panels shown, 
we have set $\alpha = -1.5$. 

The results displayed in Fig.~\ref{fig:SignificancePanelsXe}
are fundamentally due to the interplay between the individual
$dR/dE_R$ contributions from two different classes of $\chi_j$ within a given
ensemble: those with masses $m_j \lesssim 20$~GeV (the ``low-mass dark-matter'' regime)
and those with $m_j \gtrsim 20$~GeV (the ``high-mass dark-matter'' regime). 
As illustrated in Fig.~\ref{fig:StdDMdrdERPlots},
contributions to the recoil-energy spectrum from the $\chi_j$ in the low-mass regime 
begin to fall precipitously within or below the recoil-energy window for our 
hypothetical detector.  Moreover, the value of $E_R$ at which this drop-off occurs is quite 
sensitive to $m_j$ for $\chi_j$ in this regime.  By contrast, this suppression effect 
only becomes manifest for the $\chi_j$ in the high-mass regime at $E_R$ values far beyond 
the recoil-energy-acceptance window of our detector.  The spectra fall far more gradually
with $E_R$ for these fields, and their overall shape depends far less sensitively on $m_j$. 

This distinction between these two mass regimes plays a critical role in    
determining the significance with which one can distinguish DDM ensembles from 
traditional dark-matter candidates at direct-detection experiments.
For example, it implies that any DDM ensemble in which
all of the constituents in the ensemble fall within the high-mass regime is 
generically difficult to distinguish from traditional dark-matter candidates which 
likewise fall in the high-mass regime.  Indeed, we see in 
Fig.~\ref{fig:SignificancePanelsXe} that it is quite difficult to distinguish 
a DDM ensemble in situations in which $m_0 \gtrsim 20$~GeV, irrespective of
the values of $\beta$ and  $\delta$.  A similar behavior is also manifest in 
the region of parameter space within which $m_0 \lesssim 5$~GeV 
and $\Delta m \gtrsim 20$~GeV, again regardless of $\beta$ and  $\delta$.  This 
arises because the vast majority of nuclear recoils initiated by 
any $\chi_j$ with $m_j \lesssim 5$~GeV have $E_R$ values which fall below the 
detector threshold $\Emin$.  The contribution to the total recoil-energy spectrum 
for any $\chi_j$ with $m_j$ in this region is therefore essentially invisible.  
Consequently, in cases in which $m_0 \lesssim 5$~GeV while $\Delta m\gtrsim 20$~GeV, 
only the contributions from the $\chi_j$ in the high-mass regime are evident 
and the distinguishing power is once again low.      

By contrast, within other substantial regions of the parameter space of our simplified
DDM model, the statistical significance of differentiation is quite high. 
For example, in each panel displayed in Fig.~\ref{fig:SignificancePanelsXe}, 
there exists a particular range of $\Delta m$ values within which a $5\sigma$
significance is obtained for $5\lesssim m_0 \lesssim 20$~GeV.
Within this region, the kink behavior evinced in several of the recoil-energy 
spectra displayed in Fig.~\ref{fig:dRdERPanelsXe} can be distinguished.
The range of $\Delta m$ values within which this is possible depends
primarily on the value of $\beta$.  When $\beta$ is small and the coupling to the 
heavier $\chi_j$ is suppressed, the prospects for distinguishing a DDM ensemble on
the basis of this feature becomes significant when $\Delta m$ is such that the 
mass of the next-to-lightest constituent $\chi_1$ lies just above the 
threshold $m_1 \sim 20$~GeV of the high-mass regime.  These prospects remain high
until $\Delta m$ reaches the point at 
which the collective contribution to the recoil-energy spectrum from the $\chi_j$ 
in the high-mass regime falls below the sensitivity of the detector.     
As $\beta$ increases, this contribution remains substantial for larger and larger 
$\Delta m$.  However, increasing $\beta$ also results in this contribution becoming 
sufficiently large for small $\Delta m$ that it overwhelms the 
contribution from $\chi_0$ and yields an overall spectrum
indistinguishable from that of a traditional dark-matter candidate with $m_\chi$
in the high-mass regime.  The consequences of these two effects are apparent in
Fig.~\ref{fig:SignificancePanelsXe}, which shows how the region of elevated significance
due to the resolution of a kink in the recoil-energy spectrum shifts from 
$10\mathrm{~GeV}\lesssim \Delta m \lesssim 50\mathrm{~GeV}$ for $\beta = -2$ to
approximately $70\mathrm{~GeV}\lesssim \Delta m \lesssim 800\mathrm{~GeV}$ 
for $\beta = 0$.  
  
Another region of parameter space within which kinks in the 
recoil-energy spectrum frequently lead to an enhancement in the significance of 
differentiation is that within which $m_0 \lesssim 5$~GeV and 
$7\mathrm{~GeV}\lesssim \Delta m \lesssim 20\mathrm{~GeV}$.  Indeed, such 
an enhancement is evident in many of the panels in Fig.~\ref{fig:SignificancePanelsXe}.
Within this region, $m_0$ is sufficiently light that the contribution from $\chi_0$ 
to the recoil-energy spectrum is hidden beneath the detector threshold, $m_1$ lies 
within the low-mass region, and all of the remaining $m_j$ with $j \geq 2$ lie 
within the high-mass regime.  Thus, within this region, $\chi_1$ plays the same 
role which $\chi_0$ plays in the region of parameter space discussed above. 

In a number of the panels displayed in Fig.~\ref{fig:SignificancePanelsXe} --- and
especially those in which $\delta \lesssim 1$ --- we 
obtain a sizeable significance of differentiation for our DDM ensemble not merely 
within this region, but over a substantial region of the parameter space within 
which $m_0,\Delta m \lesssim 20$~GeV.  Indeed, throughout 
much of this region, there exist multiple $\chi_j$ with closely-spaced $m_j$ in the 
low-mass regime.  
When this is the case, the corresponding recoil-energy spectrum for the DDM ensemble 
assumes the characteristic ogee shape discussed in Sect.~\ref{sec:ScatteringDDM}.
This ogee shape is a distinctive feature of DDM scenarios with small $\Delta m$, 
and serves as an effective discriminant between such scenarios and traditional 
dark-matter models.

As is evident from the results shown in Fig.~\ref{fig:SignificancePanelsXe},
the significance of differentiation depends on $\delta$ in a 
somewhat complicated manner. 
Broadly speaking, the significance of differentiation obtained for 
$m_0,\Delta m \lesssim 20$~GeV tends to decrease as $\delta$ decreases, 
especially for large $\beta$.  The primary reason for this is that the density of 
states in the ensemble increases rapidly with $m_j$ when $\delta$ is small, 
and thus a greater proportion of $\Omegatot$ is carried by the $\chi_j$ 
in the high-mass regime.  Provided that $\beta$ is sufficiently large that a 
sizeable number of these $\chi_j$ couple to nucleons with reasonable 
strength, their collective contribution to the differential event rate tends to 
overwhelm that of the $\chi_j$ in the low-mass regime.  Moreover, even when  
the low-mass constituents do contribute significantly to the overall rate,
the curvature of the ogee shape becomes less pronounced --- and therefore more 
difficult to distinguish --- as $\beta$ increases.  By contrast, when $\delta$
is large, the lighter $\chi_j$ carry a greater proportion of $\Omegatot$, and their
contributions to the recoil-energy spectrum are more readily resolved.  However,
increasing $\delta$ also increases the mass splittings among the lighter 
$\chi_j$.  This has the dual effect of pushing a greater and greater number of the 
$\chi_j$ into the high-mass regime and making the individual contributions 
of the remaining constituents in the low-mass region easier to resolve.  
As a result, the broad regions of parameter space throughout which a 
DDM ensemble could be distinguished on the basis of a characteristic 
ogee feature in the recoil-energy spectrum for small $\delta$ are replaced 
at large $\delta$ by a set of ``islands'' in which a kink in the spectrum 
is the distinguishing feature.  These effects are already apparent in the right 
column of Fig.~\ref{fig:SignificancePanelsXe}.

\begin{figure}[ht!]
\begin{center}
  \epsfxsize 2.25 truein \epsfbox
    {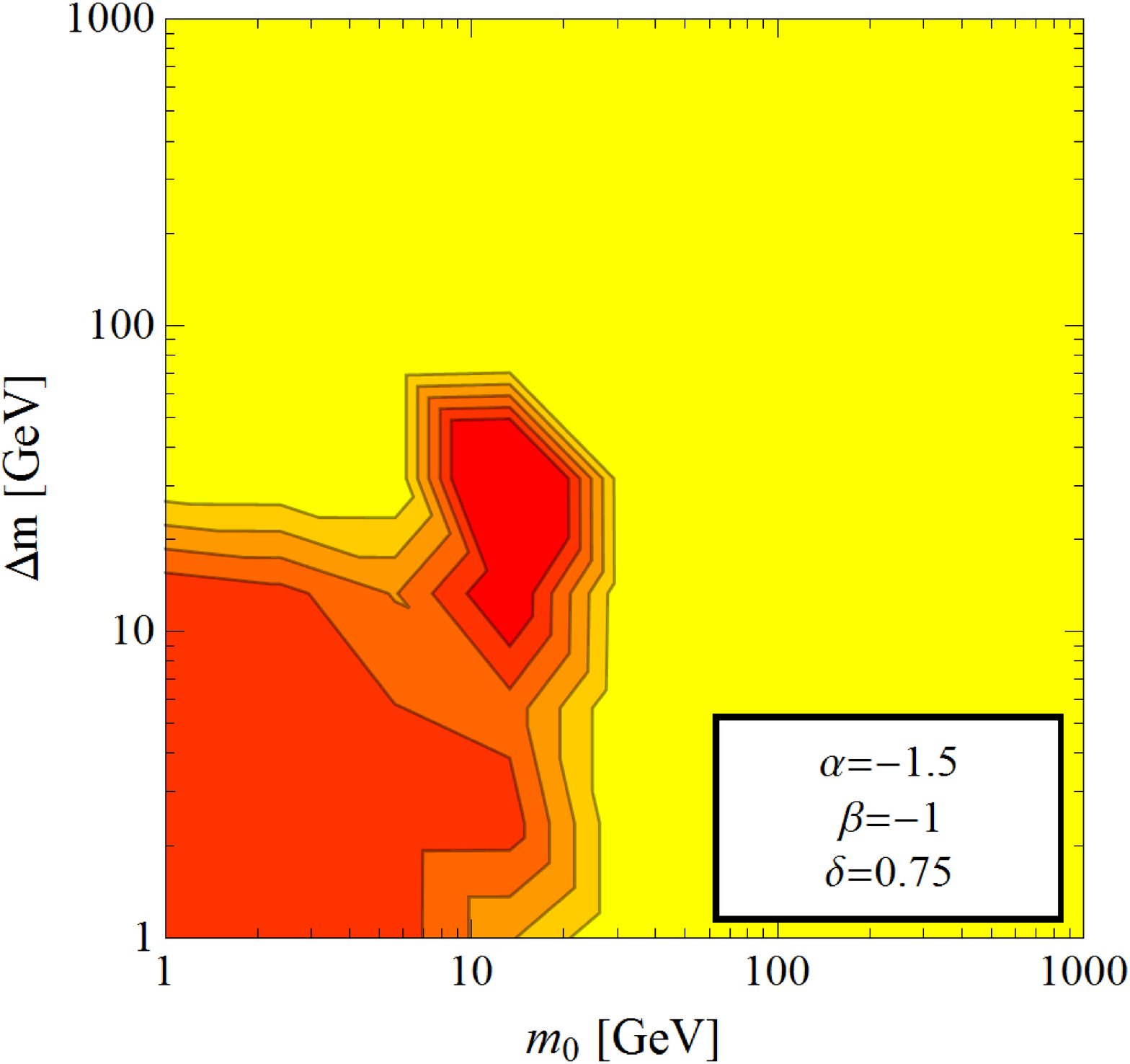}
  \epsfxsize 2.25 truein \epsfbox 
    {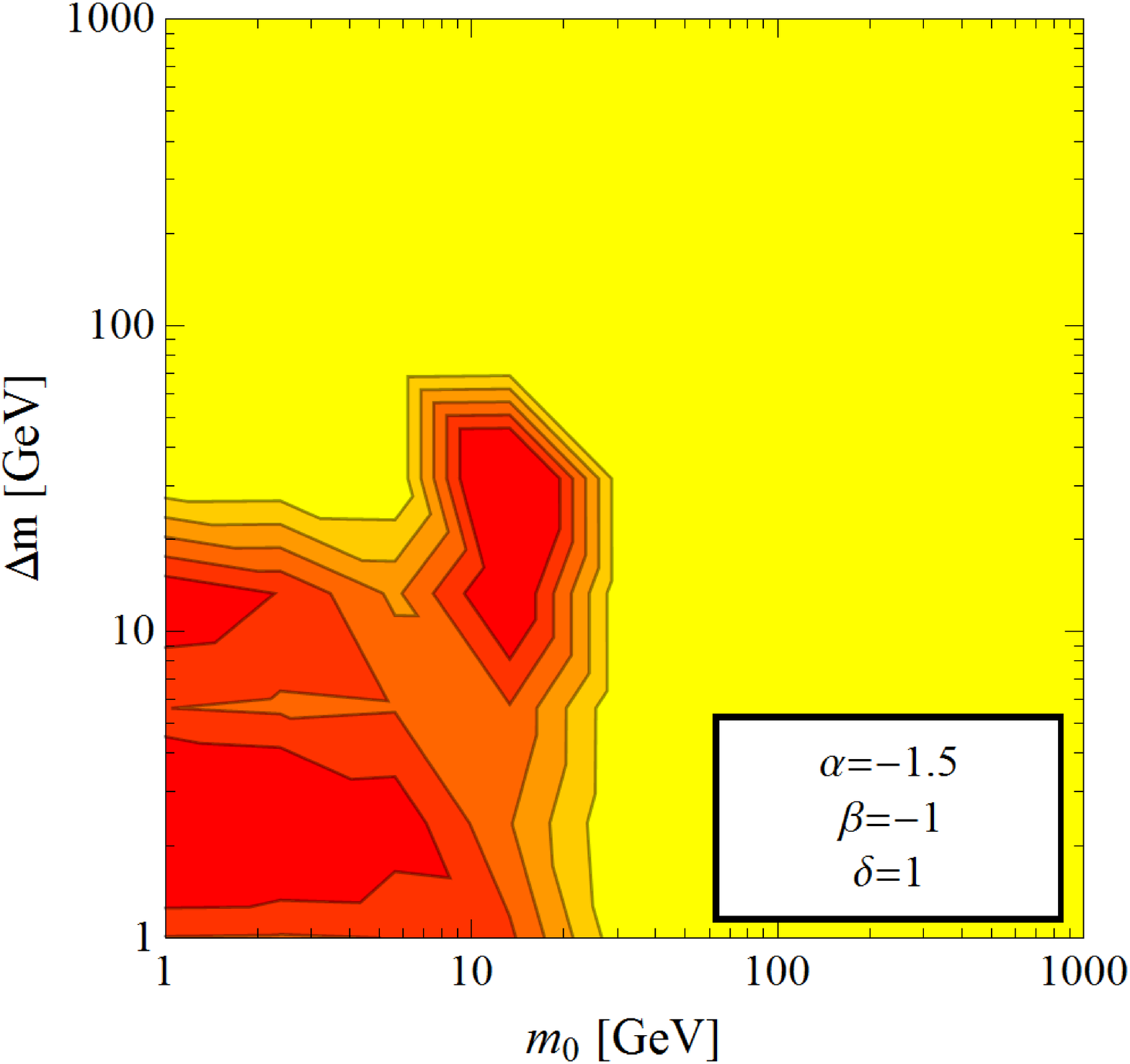}
  \epsfxsize 2.25 truein \epsfbox 
    {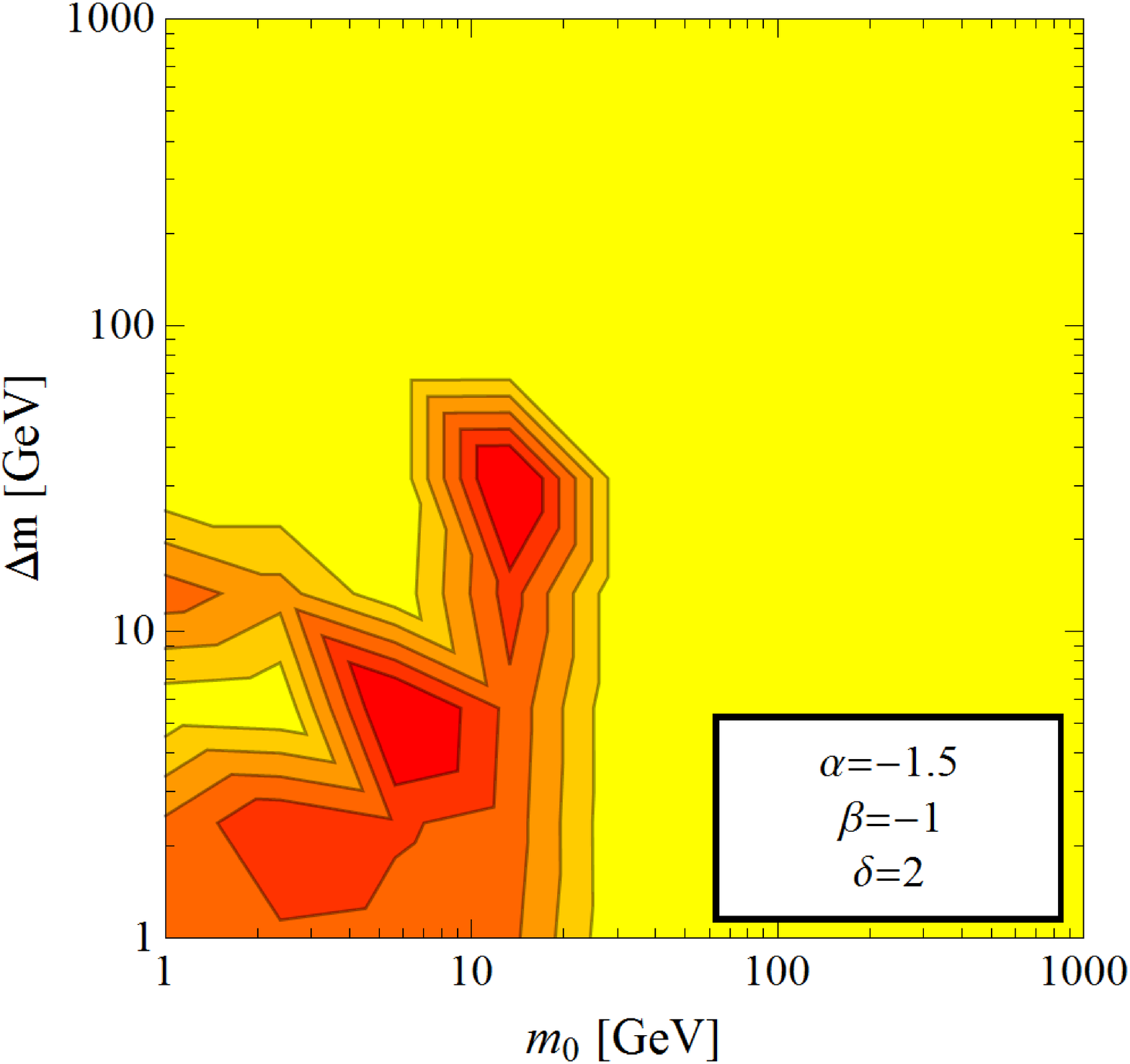}
\end{center}
\caption{
Contour plots showing the significance level at which the recoil-energy
spectrum associated with a DDM ensemble can be distinguished from that 
associated with {\it any}\/ traditional dark-matter candidate 
which gives rise to the same total event rate at a hypothetical 
germanium-crystal detector.  The colored regions shown correspond to the same 
significance intervals as in Fig.~\protect\ref{fig:dRdERPanelsXe}.  
The detector is taken to have 
a fiducial volume of 5000~kg and characteristics otherwise similar to those of 
the proposed SNOLAB phase of the SuperCDMS experiment, as discussed in the text.  
Once again, running time of five live years and an event count of $N_e=1000$ signal 
events is assumed.        
\label{fig:SignificancePanelsGe}}
\end{figure}

Let us now compare these results to the results we obtain for our hypothetical 
germanium-crystal detector.  
In Fig.~\ref{fig:SignificancePanelsGe}, we show the projected statistical 
significance of differentiation obtained with $N_e = 1000$ signal events at 
this hypothetical detector.
The panels in the left, center, and right columns of the figure correspond 
respectively to $\delta = \{0.75,1,2\}$, and in each of these three panels 
we have set $\alpha = -1.5$ and $\beta = -1$.
It is evident from Fig.~\ref{fig:SignificancePanelsGe} that 
while the quantitative results obtained for our two hypothetical detectors differ 
due to differences in the mass of the target nucleus, the observed spectrum of 
background events, \etc, the qualitative results are quite similar.  For 
$5\mathrm{~GeV}\lesssim m_0\lesssim 20\mathrm{~GeV}$, there exists a range of 
$\Delta m$ values within which the presence of a discernible kink in the 
recoil-energy spectrum leads to a $5\sigma$ significance of differentiation.
In addition, a similarly high significance is obtained for $\Delta m, m_0 \lesssim 20$~GeV 
due either to the similar kink features (at large $\delta$) or to the 
characteristic ogee shape to which DDM ensembles can give rise when mass splittings are
small (at small $\delta$).
Note that the particular significance values displayed in 
Figs.~\ref{fig:SignificancePanelsXe} and~\ref{fig:SignificancePanelsGe} 
depend on the recoil-energy threshold adopted for our hypothetical detectors 
and on how threshold effects are incorporated into the analysis, especially for 
small $m_0$.  However, varying such assumptions does not affect the qualitative 
results of our analysis.


\section{Discussion and Conclusions\label{sec:Conclusions}}


In this paper, we have investigated the potential for discovering a DDM ensemble
and differentiating it from a traditional dark-matter candidate at the next 
generation of dark-matter direct-detection experiments.  In particular, we have
assessed the degree to which these two classes of dark-matter candidates may be
distinguished on the basis of differences in recoil-energy spectra.
We have demonstrated that DDM ensembles give rise to a number of 
characteristic features in such spectra, including observable kinks and 
distinctive ogee profiles.  Moreover, we have demonstrated that under standard 
assumptions, the identification of such features can serve to distinguish a DDM 
ensemble from any traditional dark-matter candidate at the $5\sigma$ 
significance level at the next generation of direct-detection experiments.  We have
found that the prospects for differentiation are particularly auspicious in cases 
in which the mass splittings between the constituent fields in the DDM ensemble 
are small and in which the mass of the lightest such field is also relatively small.
Note that this is also a regime in which a large fraction of the full DDM ensemble 
contributes meaningfully to $\OmegaCDM$.

It is also interesting to compare the prospects for distinguishing
DDM ensembles at direct-detection experiments to the 
prospects for distinguishing them at the LHC.  We have demonstrated here
that the former are greatest when $m_0 \lesssim 20$~GeV, $\Delta m$ is small,
$0.25 \lesssim\delta \lesssim 2$, and the effective couplings between the $\chi_j$ 
and SM particles decrease moderately with $m_j$.  By contrast, it 
was shown in Ref.~\cite{DDMColliders} that the latter are greatest when 
$\Delta m$ is small, $\delta \lesssim 1$, and the effective couplings 
to SM particles increase with $m_j$.  Thus, we see that these two experimental
methods of distinguishing DDM ensembles are effective in somewhat different 
regions of parameter space, and are therefore complementary.  However, we note
that there is one region in which evidence for a DDM ensemble may 
manifest itself both at direct-detection experiments and at the LHC.  This is the
region in which $m_0 \lesssim 20$~GeV, $0.25 \lesssim \delta \lesssim 0.75$, 
the effective couplings to SM particles are roughly independent of mass, and
$\Delta m$ is also either quite small or else within the range in which an 
observable kink arises in the recoil-energy spectrum.  The simultaneous 
observation of both collider and direct-detection signatures in this case 
would provide highly compelling experimental evidence for a DDM ensemble.  

Needless to say, there are numerous additional directions potentially relevant 
for direct detection which we have not explored in this paper.  For example,
we have not considered the prospects for distinguishing DDM ensembles at 
argon- or carbon-based detectors, or instruments involving 
target materials other than xenon and germanium.  Likewise,
we have not considered the prospects for observing an annual 
modulation in the signal rate at direct-detection experiments --- a strategy
long employed by the DAMA experiment and more recently by CoGeNT.  We have also 
not considered directional detection.  More generally,
we have not considered modifications of the astrophysical assumptions (such 
as the halo-velocity distributions) or nuclear-form-factor model which define 
our standard benchmark.  Finally, we have not endeavored to 
compare or correlate signals from multiple detectors using different target 
materials.  These directions are all ripe for further study~\cite{FutureDDMDDPaper}.  

In a similar vein, in this paper we have restricted our attention to cases 
in which elastic processes dominate the scattering rate for all particles in the 
DDM ensemble.  However, within the context of the DDM framework,
inelastic scattering 
processes~\cite{InelasticDM,TuckerSmithInelasticDM,MagneticFluffy} 
of the form 
$\chi_j N\rightarrow \chi_k N$ where $j\neq k$ also occur, and can contribute 
significantly to this rate when $\Delta m \lesssim \mathcal{O}(100\mathrm{~keV})$.
This possibility is particularly interesting for a number of reasons. 
For example, in DDM scenarios, the final-state particle in such inelastic 
scattering events can be a heavier particle in the ensemble, 
as in typical inelastic dark-matter models, but it can also be a {\it lighter} 
particle in the ensemble.  In other words, inelastic scattering in the DDM 
framework involves both ``upscattering'' and ``downscattering'' processes.
This latter possibility is a 
unique feature of DDM scenarios, given that the initial-state particle $\chi_j$ 
need not be the lightest particle in the dark sector. 
Moreover, as we have 
demonstrated above, the range of $\Delta m$ relevant for inelastic scattering 
is also one in which the characteristic features to which DDM ensembles give 
rise are particularly pronounced.   

Some of the consequences of such inelastic processes are readily apparent.
For example, let us consider the case in which 
$|\delta m_{jk}| \ll m_j,m_k$, where $\delta m_{jk} \equiv m_k - m_j$. 
Although the matrix element for inelastic 
scattering is, to leading order, of the same form as for an elastic interaction, 
the kinematics can be very different.  In the limit 
$|\delta m_{jk}| \ll m_j,m_k$, the maximum recoil energy
$E_{jk}^+$ and minimum recoil energy $E_{jk}^-$  
possible in inelastic scattering are given by
\begin{equation}
  E_{jk}^{\pm} ~=~ \frac{2 \mu_{Nj}^2 v^2}{m_N} 
    \left(1 - \frac{\delta m_{jk}}{\mu_{Nj} v^2}
    \pm\sqrt{1 - 2\frac{\delta m_{jk}}{\mu_{Nj} v^2}} \right)
\end{equation}
where $v$ is the relative velocity of the initial particles.  If 
$\delta m_{jk} >0$, then this upscattering process is similar to 
that typically considered in models of inelastic dark 
matter~\cite{InelasticDM,TuckerSmithInelasticDM}, 
and its basic effect is to narrow the range 
of recoil energies for which scattering is possible for a fixed dark-matter 
velocity relative to the Earth.  A general result of this effect is that 
heavier components of the DDM ensemble are ``brought into range" of 
a direct-detection experiment, which can then resolve the recoil-energy 
endpoint.  For $\delta m_{jk} <0$, however, the range of possible 
recoil energies is broadened.  As a result, low-mass members of the DDM 
ensemble can produce recoils which lie above the recoil-energy 
threshold $E_R^{\mathrm{min}}$ of a particular experiment.  
Moreover, we note that the 
matrix element for the process $\chi_j N \rightarrow \chi_k N$ 
determines the matrix element for the process $\chi_k N \rightarrow \chi_j N$ 
through crossing symmetry.  For a DDM ensemble with a fixed distribution of 
densities, the scattering rates of different components can thus be related 
to each other.  All of these possibilities will be discussed 
further in Ref.~\cite{DDMDDInelastic}.


\begin{acknowledgments}


We would like to thank J.~Cooley, E.~Edkins, K.~Gibson, J.~Maricic, 
and J.~T.~White for discussions. 
JK and BT would also like to thank the Center for Theoretical Underground Physics 
and Related Areas (CETUP$^\ast$ 2012) in South Dakota for its hospitality and for 
partial support during the completion of this work.
KRD is supported in part by the U.S. Department of Energy under 
Grant No.\ DE-FG02-04ER-41298 and by the National 
Science Foundation through its employee IR/D program.
JK and BT are supported in part by DOE Grant No.\ DE-FG02-04ER-41291.  The opinions and 
conclusions expressed herein are those of the authors, and do not represent either 
the Department of Energy or the National Science Foundation.

\end{acknowledgments}


\end{document}